\documentclass[fleqn,usenatbib,twocolumn]{mnras}

\usepackage[T1]{fontenc}
\usepackage{ae,aecompl}

\usepackage{graphicx}	% Including figure files
\usepackage{amsmath}	% Advanced maths commands
\usepackage{amssymb}	% Extra maths symbols
\usepackage{hyperref}
\usepackage[utf8]{inputenc}
\usepackage{mathrsfs}
\usepackage{bm}
\usepackage{hyperref}
\usepackage[dvipsnames]{xcolor}

\newcommand{\be}{\begin{equation}}	
\newcommand{\ee}{\end{equation}}	
		
\newcommand{\refig}[1]{Fig.~\ref{#1}}
\newcommand{\refeq}[1]{Eq.~\ref{#1}}

\newcommand{\hydro}{H}
\newcommand{\aligneddip}{L1-0}
\newcommand{\tilteddip}{L1-90}
\newcommand{\quadA}{L2-0A}  
\newcommand{\quadB}{L2-0B}

\newcommand{\hydrotwod}{H\_2d}
\newcommand{\aligneddiptwod}{L1-0\_2d}
\newcommand{\quadAtwod}{L2-0A\_2d}
\newcommand{\quadBtwod}{L2-0B\_2d}
\title[3D CCSN with complex magnetic structures]{Three-dimensional core-collapse supernovae with complex magnetic structures: I. Explosion dynamics}

\author[M. Bugli et al.]{
M Bugli,$^{1}$\thanks{E-mail: matteo.bugli@cea.fr}
J. Guilet,$^{1}$
and M. Obergaulinger$^{2,3}$
\\
$^{1}$Laboratoire AIM, CEA/DRF-CNRS-Universit\'e Paris Diderot, IRFU/D\'epartement d'Astrophysique, CEA-Saclay F-91191, France\\
$^{2}$Departament d'Astronomia i Astrof\'isica, Universitat de Val\`encia, Dr. Moliner 50, 46100, Burjassot, Spain\\
$^{3}$Institut f{\"u}r Kernphysik, Theoriezentrum, Schlossgartenstr. 2, D-64289 Darmstadt, Germany
}

\date{Accepted XXX. Received YYY; in original form ZZZ}

\pubyear{2021}

\usepackage{newtxtext,newtxmath}

\begin{document}
\label{firstpage}
\pagerange{\pageref{firstpage}--\pageref{lastpage}}
\maketitle

\begin{abstract}
Magnetic fields can play a major role in the dynamics of outstanding explosions associated to violent events such as GRBs and hypernovae, since they provide a natural mechanism to harness the rotational energy of the central proto-neutron star and power relativistic jets through the stellar progenitor.
As the structure of such fields is quite uncertain, most numerical models of MHD-driven core-collapse supernovae consider an aligned dipole as initial magnetic field, while the field's morphology can actually be much more complex.
We present three-dimensional simulations of core-collapse supernovae with more realistic magnetic structures, such as quadrupolar fields and, for the first time, an equatorial dipolar field.
Configurations other than an aligned dipole produce weaker explosions and less collimated outflows, but can at the same time be more efficient in extracting the rotational energy from the PNS.
This energy is then stored in the surroundings of the PNS, rather than powering the polar jets.
A significant axial dipolar component is also produced by models starting with a quadrupolar field, pointing to an effective dynamo mechanism operating in proximity of the PNS surface.

\end{abstract}


%%%%%%%%%%%%%%%%%%%%%%%%%%%%%%%%%%%%%%%%%%%%%%%%%%

%%%%%%%%%%%%%%%%% BODY OF PAPER %%%%%%%%%%%%%%%%%%

\section{Introduction}
The gravitational collapse of a massive star is one of the most violent events occurring in the Universe, as it releases a huge amount of gravitational binding energy (of the order of $10^{53}$ erg) within just a few seconds.
While most of this energy (about 99\%) is carried away by neutrinos during the formation and cooling of the proto-neutron star (PNS) at the centre of the stellar progenitor, the remaining fraction can be enough to power the outward propagation of the shock wave formed at core bounce and lead to a core-collapse supernova explosion (CCSN).
The vast majority of these events rely on the so-called neutrino-heating mechanism to launch a successful explosion \citep[for a review, see, e.g.,][]{Janka2012}, where a fraction of the neutrinos emitted from the PNS deposit their energy below the shock wave and enhance the thermal pressure.
Although such a scenario can quantitatively account for the properties of most observed CCSN lightcurves, it cannot explain the exceptionally high luminosities of superluminous supernovae \citep{nicholl2013,greiner2015} unless considering the formation of strong shocks produced by the interaction with dense circumstellar material \citep{smith2014,inserra2017}.
Moreover, the neutrino-heating mechanism leads to the production of ejecta whose kinetic energy falls one order of magnitude short of the values inferred in outstanding supernova explosions such as hypernovae \citep{iwamoto1998,soderberg2006,drout2011}.

Magnetic fields are a promising candidate to explain the extraordinary energy budget of such outstanding transients, as they provide an efficient way to extract rotational energy from the central PNS via magnetic braking and power the launch of so-called magnetorotational explosions.
Although strong magnetic fields and fast rotation are both fundamental ingredients of this mechanism, it is still a matter of debate how it could be possible to produce a combination of the two during the collapse of a massive star.
Magnetic fields can be amplified by convective dynamos during the life of the stellar progenitor, but it is not clear whether this scenario could produce magnetic fields that would lead to magnetar-like values of $\sim 10^{15}$ G after the gravitational collapse.
A physical process like the Tayler instability \citep{spruit2002} could explain the field amplification in the stably stratified layers of the progenitor, but this mechanism tends also to increase the transport of angular momentum from the core to the envelope and enhance its losses via magnetically driven winds, thus slowing down the star's rotation \citep{ma2019}.
Considering the scenario in which the magnetic field is of fossil origin also leads to the conclusion that strong magnetic fields are likely to be associated to slow rotation.
The strong fields observed at the surface of Ap, Bp and Of?p stars are indeed expected to brake the surface rotation through magnetic winds \citep{shultz2018} and to brake the core by connecting it to the envelope.
Finally, the strong magnetic field generated after the merger of two main sequence stars is also expected to be associated with slow-rotation because \cite{schneider2019} found that the post-merger star had such a slow rotation.

%PNS DYNAMOS
A possible solution to this problem is the in situ amplification of a weak magnetic field during the gravitational collapse inside of the PNS through some dynamo process, which would tap into the rotational energy of the stellar progenitor to produce a strong large-scale field and enable the launch of the explosion.
Recent studies have shown that such amplification could be linked to the dynamics of the convectively unstable layers of the PNS, with a weak seed field being amplified up to magnetar-like values %within few hundreds of milliseconds} 
\citep{raynaud2020}.
Another possible dynamo process could be due to the magnetorotational instability \citep{balbus1998}, which is expected to occur in the external layers of the PNS \citep{akiyama2003,obergaulinger2009,guilet2015} and could produce strong large-scale magnetic fields from small-scale fluctuations \citep{reboul-salze2021}.

%KINK AND 3D INSTABILITIES
While the efficiency and viability of the magnetorotational explosion mechanism have been extensively investigated in the last two decades via axisymmetric simulations of magnetised CCSN \citep{akiyama2003,ardeljan2005,sawai2005,obergaulinger2006a,obergaulinger2006b,burrows2007,dessart2007,takiwaki2009,obergaulinger2014a,obergaulinger2017,obergaulinger2020a}, the same cannot be said for three-dimensional (3D) models, since a more limited number of such studies has been so far conducted \citep{scheidegger2008,mosta2014b,kuroda2020,obergaulinger2021a} due to their high computational cost.
Given the large number of parameters that can affect the dynamics of magnetorotational explosions and the relatively high computational cost of 3D simulations, it is still not clear to what extent the results of axisymmetric models may still be valid when increasing the dimensionality of the problem.
Physical processes such as convection and amplification of magnetic fields by dynamo action are inherently 3D, and can have a deep qualitative impact on the system's dynamics.

One of the most important open questions currently investigated is to what extent is the explosion affected by the growth of non-axisymmetric instabilities, which are obviously filtered out in 2D models.
One prominent example is the so-called \emph{kink instability}, which leads to the excitation of large-scale azimuthal modes in the expanding ejecta whose barycentre gets consequently displaced from the rotational axis.
Some studies have shown that the growth of the kink instability can significantly disrupt the coherence of the polar outflows, possibly even preventing a successful explosion \citep{mosta2014b,kuroda2020}.
However, \cite{obergaulinger2021a} recently presented a series of models that, although subject to the kink instability, launched successful magnetorotational explosions, and in some cases proved to be qualitatively similar to their axisymmetric counterparts.
Another important phenomenon is the onset of corotational instabilities in the PNS \citep{passamonti2015}, which requires a sufficiently fast rotation in the stellar progenitor and can induce large-scale non-axisymmetric perturbations within the PNS, significantly affecting the explosion dynamics \citep{takiwaki2016} and leading to a substantial emission of gravitational waves \citep{shibagaki2020}.

%MAGNETIC CONFIGURATION
An important limitation of numerical models of magnetised CCSN is that the initial magnetic configurations employed are often chosen out of practical convenience and simplicity, rather than being justified by solid observational or theoretical constraints.
Our current knowledge of the magnetic field threading the stellar progenitor at the moment of collapse is still quite uncertain, as the one-dimensional stellar evolution models that attempt to self-consistently include the dynamics of magnetic fields \citep{woosley2006,aguilera-dena2018} have to rely on describing the stellar dynamos action with strong approximations \citep{spruit2002,fuller2019}.
If we consider the amplification of magnetic fields via convection and MRI in the PNS, the saturated magnetic field is independent of the initial magnetic seed \citep{raynaud2020,reboul-salze2021}, but it is still prohibitive to properly resolve these complex dynamics within a large-scale CCSN simulation.
For these reasons, the majority of currently published 3D models employ a simple configuration for the initial magnetic field, i.e. an aligned dipolar field which is roughly constant within a characteristic radius $r_0$ and then decays.
A notable exception is provided in \cite{halevi2018}, where such an initial field was tilted up to $45^\circ$, resulting in increasingly less collimated ejecta and slower expanding shocks.
However, this study is focused on nucleosynthesis and gives little description of the explosion dynamics and evolution of the PNS.
\cite{bugli2020} explored the evolution of axisymmetric magnetised CCSN using magnetic fields with multipolar order up to $l=4$, and found that fields distributed on smaller angular scales than a magnetic dipole are still capable of sustaining a magnetorotational explosion, although producing less energetic ejecta and slower expanding shocks.
Moreover, they showed that the PNS evolution is significantly affected, as it tends to become more massive and to spin faster with increasingly higher magnetic multipoles.

%PRESENTATION
In this work we aim at extending the findings of \cite{bugli2020} to a full 3D framework, employing the same stellar progenitor and similar magnetic configurations.
In addition, we explore also the evolution of a model threaded by an equatorial dipolar field, motivated by the results of a recent MRI-driven dynamo study that shows that highly tilted dipolar components are likely to develop during the early evolution of the PNS \citep{reboul-salze2021}.
We will focus on the impact of magnetic fields on the explosion dynamics and the evolution of the PNS, comparing the effects of models with different magnetic topology and dimensionality.
We leave the analysis of co-rotational instabilities, their interaction with magnetic fields and their influence on the emission of multimessenger signals to a following paper.
The initial setup of our models is presented in Section \ref{sec:setup}, while in Section \ref{sec:discussion} we discuss the properties of the explosion, the evolution of the PNS, the development of the kink instability and the dynamics of the magnetic field.
Finally, in Section \ref{sec:conclusions} we present our conclusions and future perspectives.

%%%%%%%%%%%%%%%%%%%%%%%%%%%
%%%%%%%%%%%%%%%%%%%%%%%%%%%
%%%%%%%%%%%%%%%%%%%%%%%%%%%
%%%%%%%%%%%%%%%%%%%%%%%%%%%
%%%%%%%%%%%%%%%%%%%%%%%%%%%
%%%%%%%%%%%%%%%%%%%%%%%%%%%
%%%%%%%%%%%%%%%%%%%%%%%%%%%
%%%%%%%%%%%%%%%%%%%%%%%%%%%
%%%%%%%%%%%%%%%%%%%%%%%%%%%
%%%%%%%%%%%%%%%%%%%%%%%%%%%
%%%%%%%%%%%%%%%%%%%%%%%%%%%
%%%%%%%%%%%%%%%%%%%%%%%%%%%

\section{Numerical setup}\label{sec:setup}

We use the same numerical setup as the 2D simulations of \cite{bugli2020}, except for the non-axisymmetric magnetic configurations described below, the resolution employed and the fact that the simulations are three-dimensional.
We therefore give only a short summary of this setup and refer to \cite{bugli2020} for more details.

\begin{figure}
    \centering
    \includegraphics[width=0.5\textwidth]{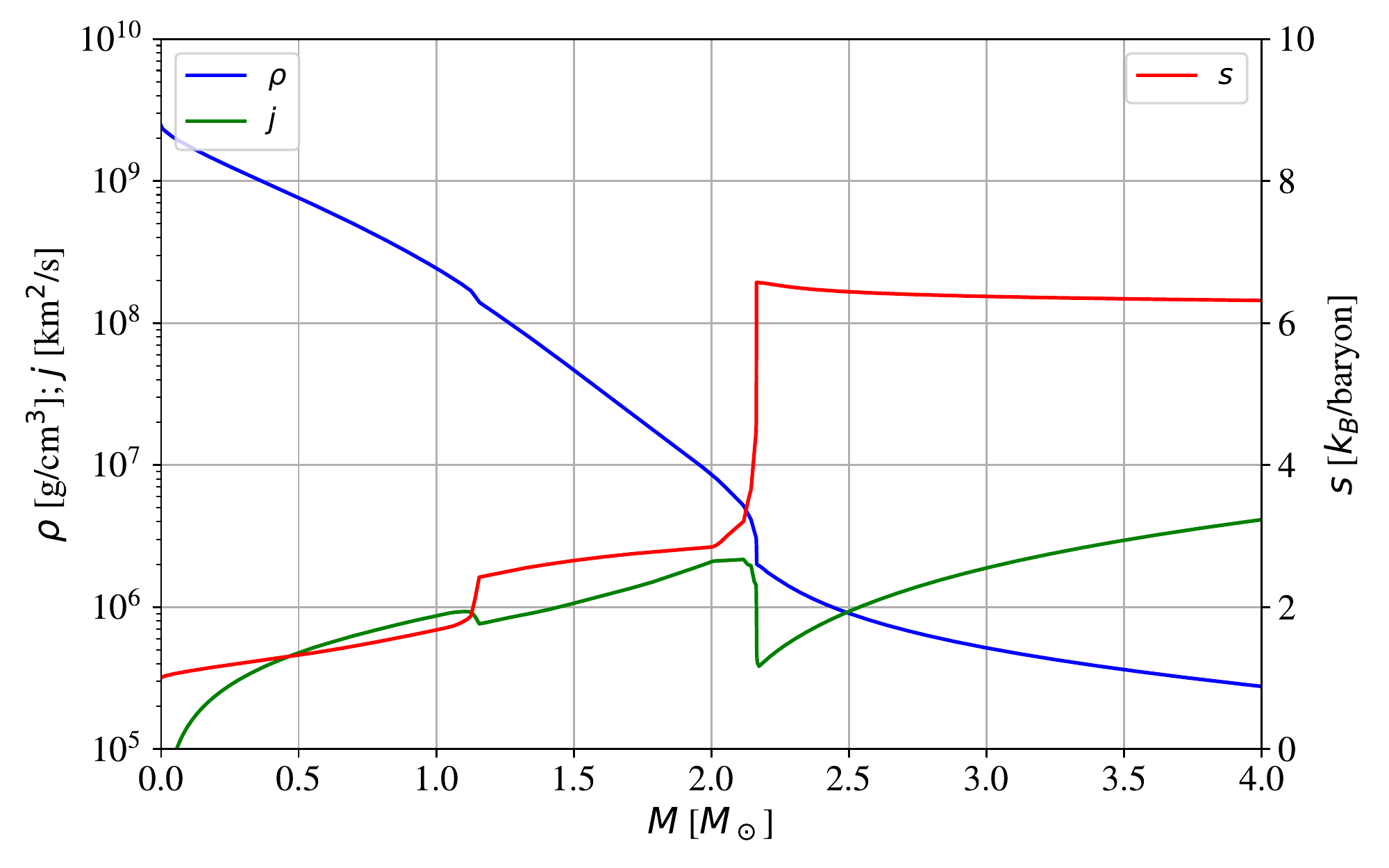}
    \caption{Profiles of mass density (blue curve), specific entropy (red) and specific angular momentum (green) of the progenitor \texttt{350C} as a function of the mass coordinate.}
    \label{fig:progenitor}
\end{figure}

The numerical models presented in this study were produced with the \texttt{AENUS-ALCAR} code \citep{obergaulinger2008,just2015}, which solves the equations of magnetohydrodynamics in special relativity coupled to a multi-group neutrino transport using an M1 scheme.
They employ the nuclear equation of state of \cite{lattimer1991} with an incompressibility of $K=220$ MeV.
All of our simulations follow the gravitational collapse of the stellar model \texttt{35OC} based on a $M_\mathrm{ZAMS}=35M_\odot$ progenitor \citep{woosley2006}, which in recent years has been employed in several studies of magnetorotational supernovae \citep{obergaulinger2017,obergaulinger2020a,bugli2020,aloy2020,obergaulinger2021a}.
The large iron core has a mass  $M_\mathrm{Fe}\approx2.1M_\odot$, a radius   $R_\mathrm{Fe}\approx2.9$ km and rotates at its centre with angular velocity $\Omega_c\approx10$ rad/s.
The core is surrounded by a convective envelope, with a sharp decrease of the specific angular momentum by a factor of $\sim 5$ at the interface between the two (see \refig{fig:progenitor}).
Concerning the relativistic corrections to the Newtonian gravitational potential, model \quadB{} uses function \emph{B} presented in \cite{marek2006}, while all the others employ function \emph{A} instead.
The two functions differ in their description of the gravitational TOV mass defining the correction to the spherical component of the gravitational potential, with case \emph{A} reducing it by a geometrical factor and case \emph{B} neglecting instead the contributions of neutrinos and gas internal energy density.
Both prescriptions performed well in the test problems presented in \citet{marek2006}, although the former is generally preferred since it allows a consistent description of the total relativistic energy.
As there are no 3D studies probing the effects of different corrections on the dynamics of magnetorotational explosions, we therefore decided to adopt in one of our models a similar but distinct gravitational potential to assess possible deviations in 3D models.

Similarly to \cite{bugli2020}, we adopt the profiles of model \texttt{35OC} only for the hydrodynamic quantities, while we superimpose ad hoc magnetic fields of different topology.
We first set the azimuthal component of the vector potential to
\begin{equation}\label{eq:Aphi}
    A^\phi_l(r,\theta)= r\frac{B_0}{(2l+1)}\frac{r_0^{3}}{r^{3}+r_0^{3}} \frac{P_{l-1}(\cos\theta)-P_{l+1}(\cos\theta)}{\sin\theta},
\end{equation}
where $l$ is the multipolar order, $B_0$ is a normalisation constant that corresponds to the strength of the magnetic field along the vertical axis divided by a factor $\sqrt{l}$, $r_0$ is the radius of the region where the strength of the field is roughly constant and $P_l$ is the Legendre polynomial of order $l$.
For all models we fix $r_0=1000$ km and $B_0=10^{12}$ G, which results in a total magnetic energy in the numerical box of $E_\mathrm{mag}\approx10^{48}$ erg (much smaller than the total rotational energy $E_\mathrm{rot}\approx4\times10^{49}$).
We consider dipolar and quadrupolar configurations by setting $l=1,2$ respectively, where the specific value of $l$ affects the strength of the magnetic field along the rotation axis but does not modify the volume-integrated magnetic energy.
With respect to \cite{bugli2020} we impose the same radial decay for fields with different multipolar orders, allowing us to factor it out as a source of deviations between different models and focus solely on the impact of the field's angular distribution.
At the same time, by keeping the same value of $B_0$ for all magnetised models we ensure that they all have the same magnetic energy budget, whereas in \cite{bugli2020} different groups of models shared the same magnetic field strength along the vertical axis.
For model \tilteddip{} we use instead the vector potential for a dipolar field tilted by an angle $\alpha$ as in \cite{halevi2018}
\begin{align}
    A_r & =  0, \\
    A_\theta & =  -r\frac{B_0}{2}\frac{r_0^{3}}{r^{3}+r_0^{3}}\sin\phi\sin\alpha,\\
    A_\phi & =  r\frac{B_0}{2}\frac{r_0^{3}}{r^{3}+r_0^{3}}(\sin\theta\cos\alpha-\cos\theta\cos\phi\sin\alpha),
\end{align}
and we set $\alpha=\pi/2$.

Our spherical numerical grid resolves the $\theta$ and $\phi$ direction respectively with 64 and 128 points, hence with a resolution of about $\Delta\theta=\Delta\phi=2.8^{\circ}$.
Along the radial direction the grid has a uniform resolution of $\Delta r=0.5$ km up to $r\sim10$ km, where the aspect ratio of the grid cell is roughly uniform in the meridional plane, i.e. $\Delta r \approx R\Delta\theta$.
Beyond this radius, the radial grid is logarithmically stretched up to $r_\mathrm{max}\approx 8.8\times10^4$ km, for a total of 210 points.
For each 3D model we performed a corresponding axisymmetric simulation (except for model \tilteddip{}, which is inherently 3D) using the same numerical grid.
This choice will benefit our analysis of the effects connected to enabling a fully 3D dynamics in the simulations, removing possible numerical sources of ambiguity. 
However, it should be noted that the 2D simulations presented in this work have a coarser angular resolution compared to those produced in \citet{bugli2020}, having therefore a central region of uniform radial resolution of about half the size.

% Models table
\begin{table}
	\centering
	\caption{Models}
	\label{tab:example_table}
	\begin{tabular}{lccccc} % four columns, alignment for each
		\hline
		\hline
		label & $N_\phi$ & $B_0$ [G] & $l$ & $\Phi$ & $\theta_B$ [$^\circ$]\\
		\hline
		\hline
		\hydrotwod{} & 1 & 0 & - & A & -\\
		\aligneddiptwod{} & 1 & $10^{12}$ & 1 & A & -\\
		\quadAtwod{} & 1 & $10^{12}$ & 2 & A & -\\
		\quadBtwod{} & 1 & $10^{12}$ & 2 & B & -\\
		\hline
		\hydro{} & 128 & 0 & - & A & -\\
		\aligneddip{} & 128 & $10^{12}$ & 1 & A & 0\\
		\tilteddip{} & 128 & $10^{12}$ & 1 & A & 90\\
		\quadA{} & 128 & $10^{12}$ & 2 & A & 0\\
		\quadB{} & 128 & $10^{12}$ & 2 & B & 0\\
		
		\hline
		\hline
	\end{tabular}
\end{table}

\begin{figure*}
    \centering
    \includegraphics[trim={0 1cm 0 0}, clip, width=0.48\textwidth]{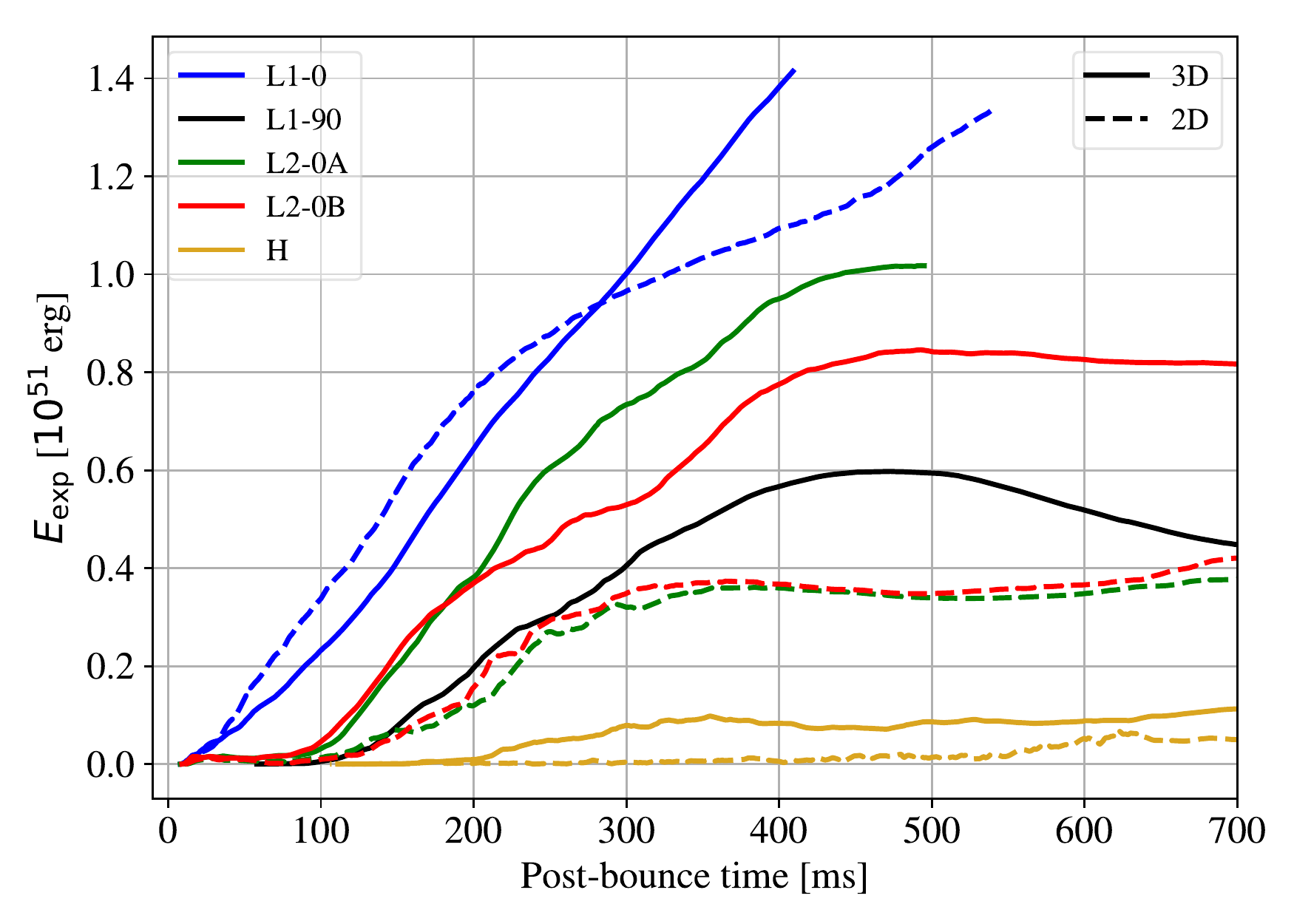}
    \includegraphics[trim={0 1cm 0 0}, clip,width=0.48\textwidth]{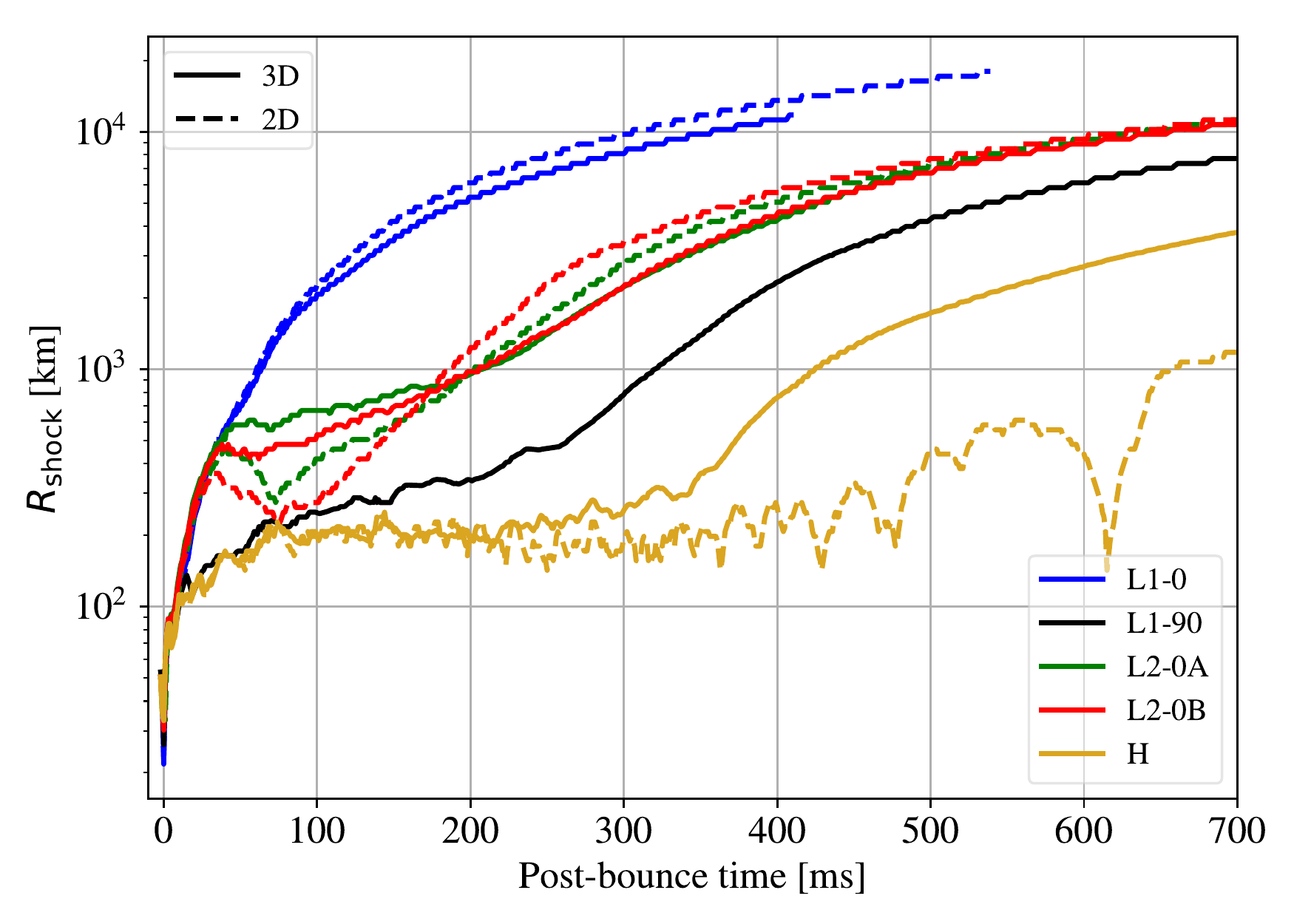}
    \includegraphics[width=0.48\textwidth]{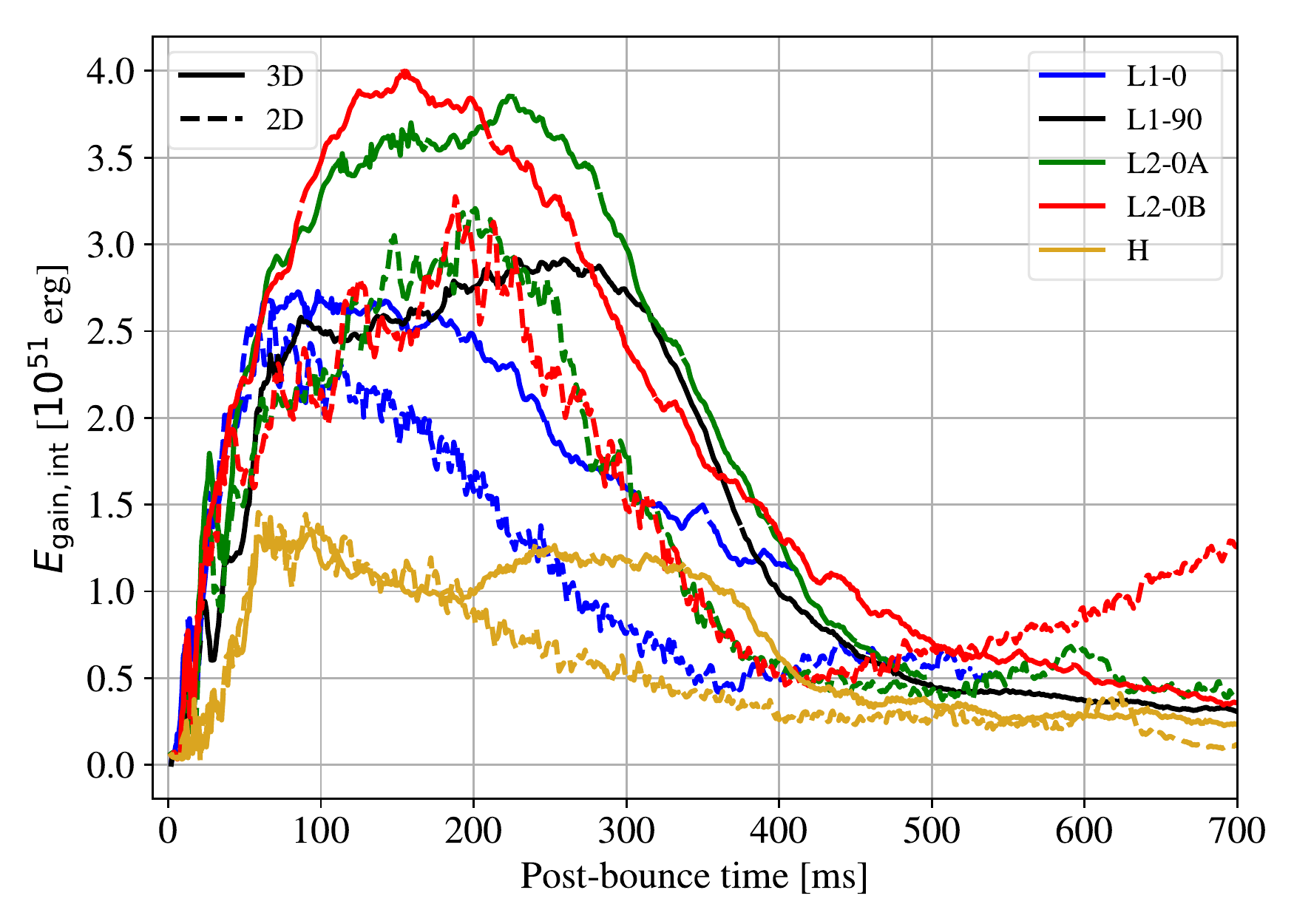}
    \includegraphics[width=0.48\textwidth]{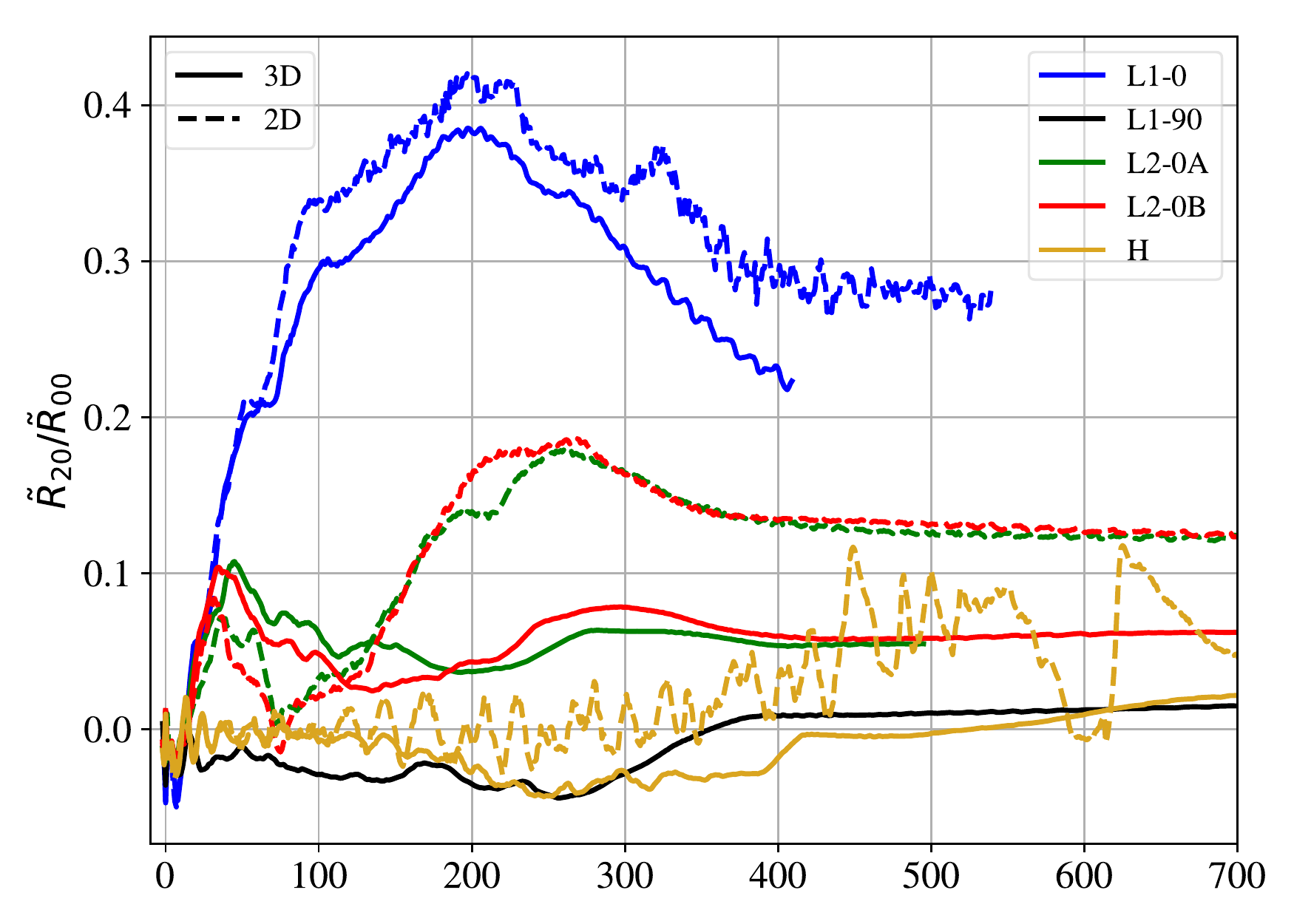}
    \caption{Time evolution of the explosion energy (top left), the shock radius along the north direction (top right), internal energy contained in the gain region (bottom left) and harmonic decomposition of the shock surface (bottom right).}
    \label{fig:expenergy_rshock}
\end{figure*}

\begin{figure*}
    \centering
    %HIGH RESOLUTION IMAGES
     \includegraphics[width=0.48\textwidth]{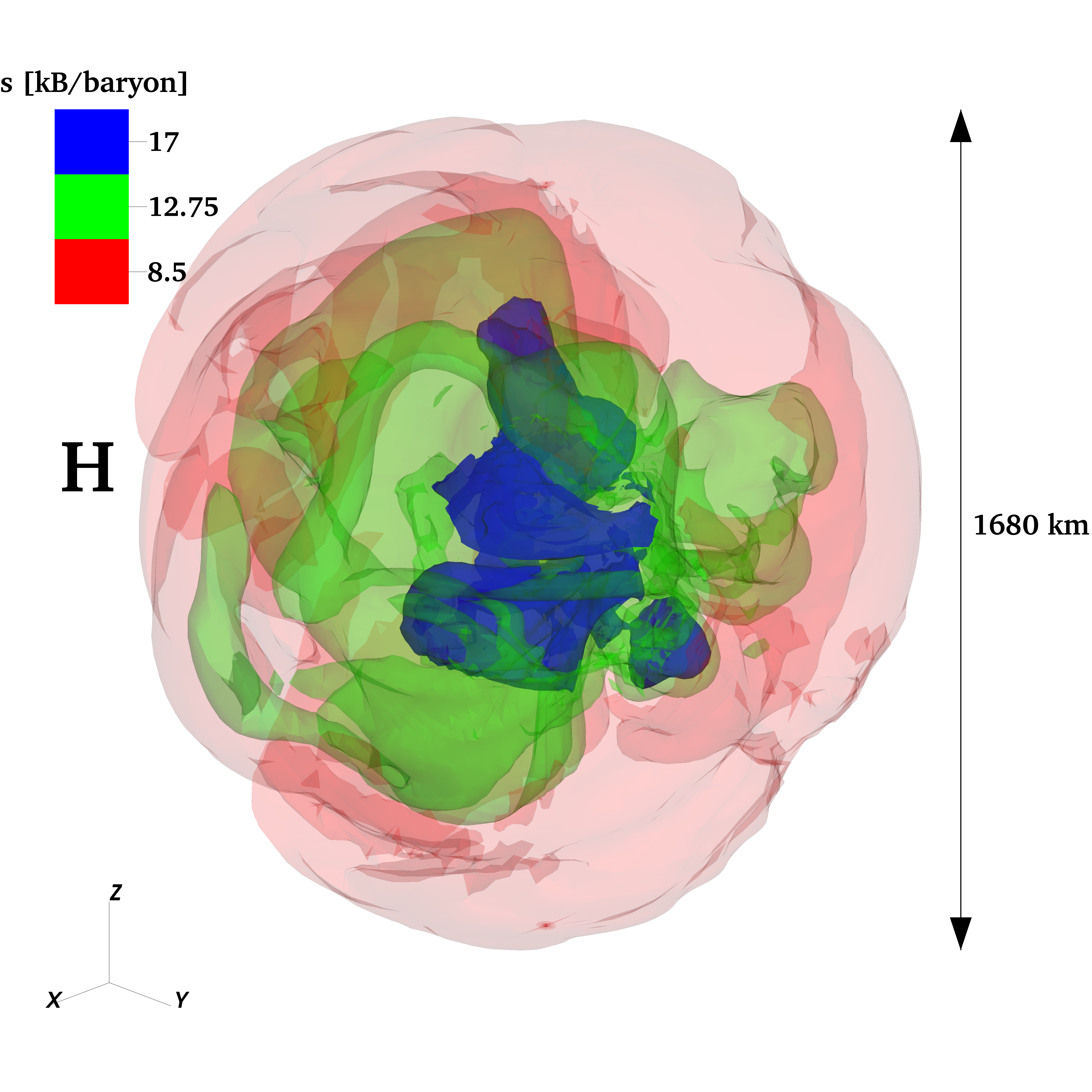}
     \includegraphics[width=0.48\textwidth]{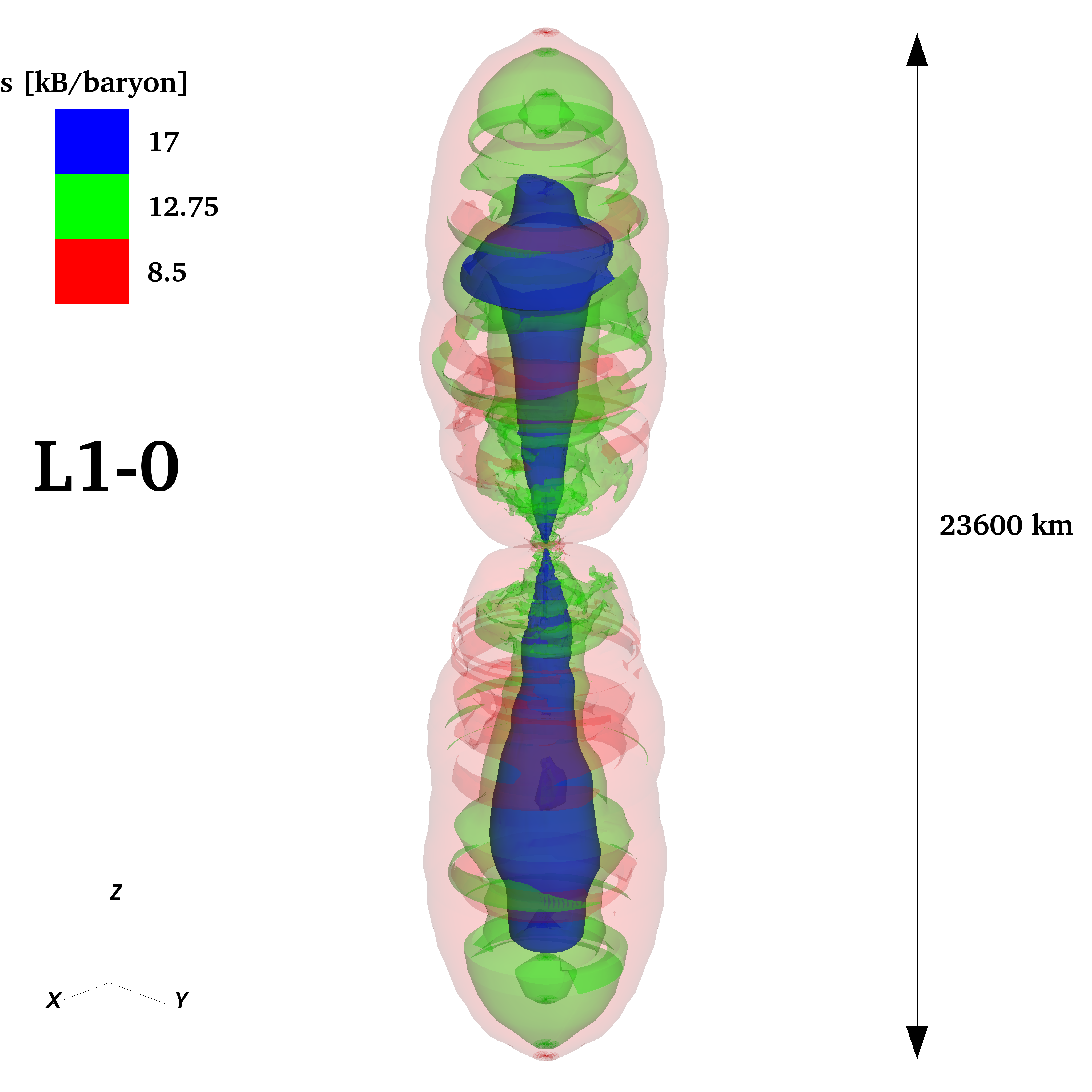}
     \includegraphics[width=0.48\textwidth]{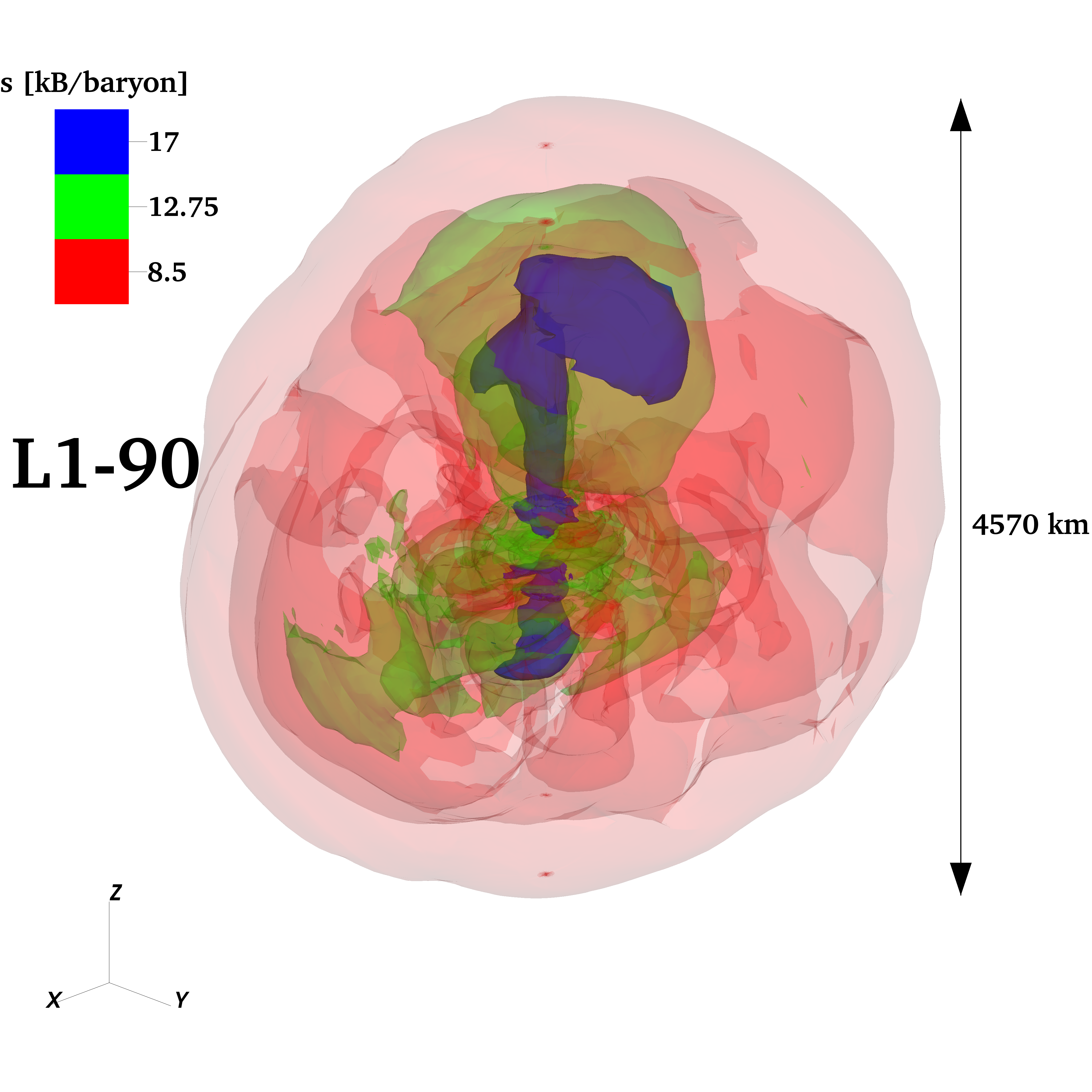}
     \includegraphics[width=0.48\textwidth]{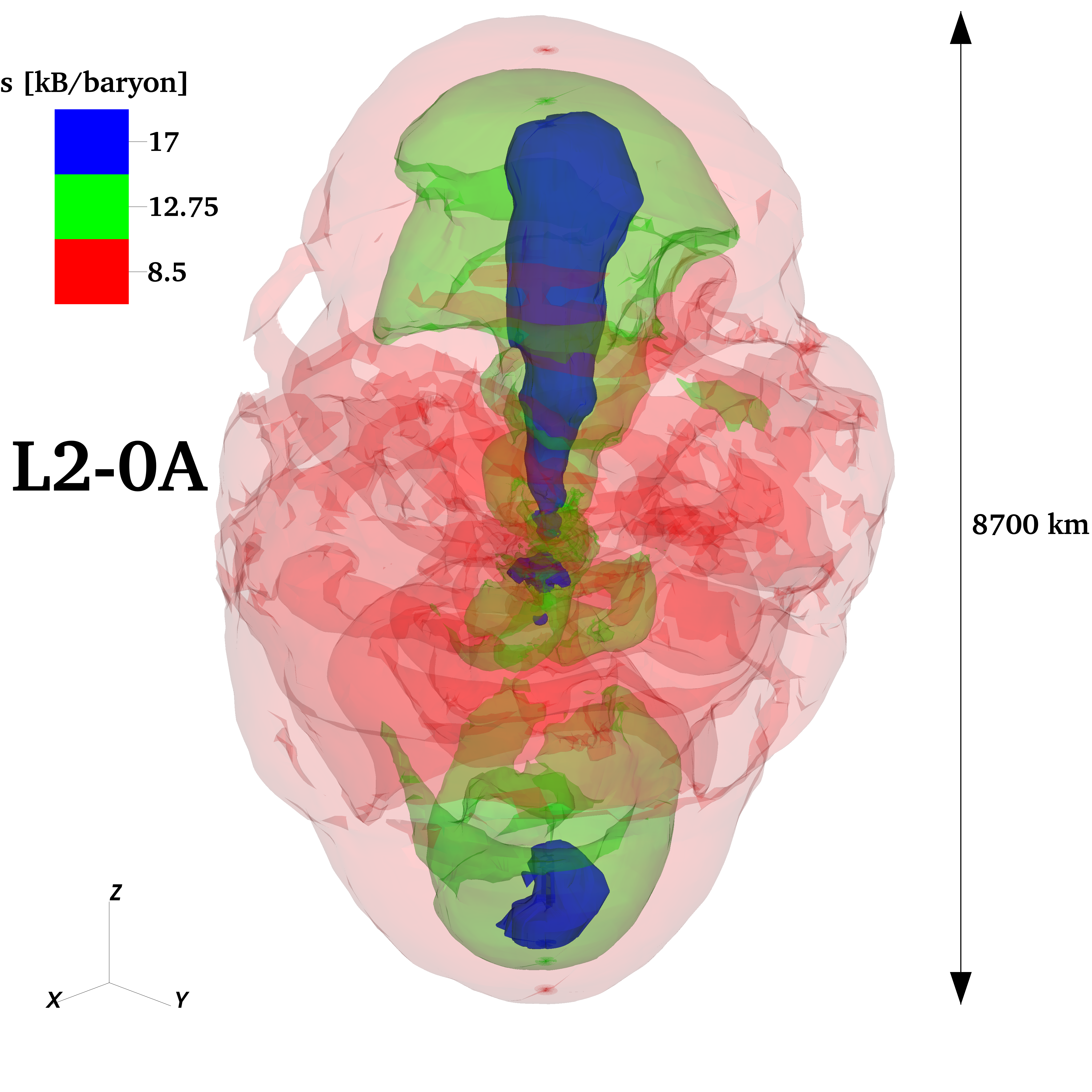}
    %LOW RESOLUTION IMAGES
    %\includegraphics[width=0.48\textwidth]{images/3d_entropy_410ms_hydro_3d-min.png}
    %\includegraphics[width=0.48\textwidth]{images/3d_entropy_410ms_l1_b12_3d-min.png}
    %\includegraphics[width=0.48\textwidth]{images/3d_entropy_410ms_l1_b12_3d_90deg-min.png}
    %\includegraphics[width=0.48\textwidth]{images/3d_entropy_410ms_l2_b12_dipdecay_3D_gravA-min.png}
    \caption{Volume rendering of the specific entropy for models \hydro{} (upper left panel), \aligneddip{} (upper right panel), \tilteddip{} (lower left panel)\ and \quadA{} (lower right panel) at $t=410$ ms p.b.}
    \label{fig:3d_entropy}
\end{figure*}

%%%%%%%%%%%%%%%%%%%%%%%%%%%
%%%%%%%%%%%%%%%%%%%%%%%%%%%
%%%%%%%%%%%%%%%%%%%%%%%%%%%
%%%%%%%%%%%%%%%%%%%%%%%%%%%
%%%%%%%%%%%%%%%%%%%%%%%%%%%
%%%%%%%%%%%%%%%%%%%%%%%%%%%
%%%%%%%%%%%%%%%%%%%%%%%%%%%
%%%%%%%%%%%%%%%%%%%%%%%%%%%
%%%%%%%%%%%%%%%%%%%%%%%%%%%
%%%%%%%%%%%%%%%%%%%%%%%%%%%
%%%%%%%%%%%%%%%%%%%%%%%%%%%
%%%%%%%%%%%%%%%%%%%%%%%%%%%
%%%%%%%%%%%%%%%%%%%%%%%%%%%
%%%%%%%%%%%%%%%%%%%%%%%%%%%
%%%%%%%%%%%%%%%%%%%%%%%%%%%
%%%%%%%%%%%%%%%%%%%%%%%%%%%
%%%%%%%%%%%%%%%%%%%%%%%%%%%
%%%%%%%%%%%%%%%%%%%%%%%%%%%

\section{Discussion}\label{sec:discussion}
\subsection{Explosion dynamics}

All of the models presented in this work produced successful explosions, regardless of the different configuration of the magnetic field.
Only in the case of an aligned dipole there is a prompt expansion of the shock, while for models with an aligned quadrupolar field the shock expands during the first 50 ms, then it stalls for about 100 ms and finally increases monotonically (see top right panel of \refig{fig:expenergy_rshock}).
Model \tilteddip{} shows a slow expansion of the shock up to $\sim250$ ms p.b., at which point the shock front begins to propagate faster away from the centre.
A similar behaviour applies to the hydrodynamic model, where the shock radius stalls until $200$ ms p.b., then expands slowly and finally begins to propagate at $\sim370$ ms p.b. at a much faster rate.
The correspondent axisymmetric models produce similar expansion histories for the shock radius, although with some notable differences.
Models \quadAtwod{} and \quadBtwod{} have shocks that initially stall at a smaller radius and then appear to expand faster than their three-dimensional counterparts.
However, after 500 ms p.b. the shock radius of models \quadA{} and \quadB{} catches up and they assume similar values to the axisymmetric ones.
In absence of magnetic fields, the shock stalls a bit longer, starting a clear expansion at about 500 ms p.b., rather than $\sim350$ ms p.b. as for model \hydro{}.

The different rates of shock expansion among various models correlate with the time evolution of the unbound ejecta diagnostic energy (see top left  panel of \refig{fig:expenergy_rshock}).
The highest energy is produced by the model with an aligned dipole, which increases monotonically and by the end of the simulation after $400$ ms p.b. reaches $\sim1.4\times10^{51}$ erg.
On the other hand, the energy of the other magnetised models peaks around $500$ ms p.b., at a value of about $\sim0.6\times10^{51}$ erg for the tilted dipole and slightly higher for the two aligned quadrupoles (respectively $10^{51}$ and $0.8\times10^{51}$ erg for models \quadA{} and \quadB{}).
Model \hydro{} produces the weakest explosion among all simulations, showing a monotonic increase of the ejecta energy for its entire duration (although quite slow, compared to the other models).

A comparison between axisymmetric and three-dimensional models clearly shows a systematically higher energy produced in three-dimensional simulations.
The only (temporary) exception seems to be the aligned dipole, for which model \aligneddiptwod{} has a slightly higher energy than \aligneddip{} until $\sim300$ ms p.b., during which time the two models show a similar growth for the ejecta energy.
However, at later times the three-dimensional model's energy takes over and keeps increasing at a faster rate.
Neutrinos play a more relevant (although still not dominant) role in our models with a quadrupolar field, with respect to model \aligneddip{}, as they do not produce a prompt magnetorotational explosion.
Instead, between 50 and 170 ms p.b. the shock front slowly increases, and it even recedes by a few tens of kilometers in the axisymmetric case.
The different behaviour between 2D and 3D models with a quadrupolar magnetic field can be explained by looking at the internal energy contained in their gain region (bottom left panel of \refig{fig:expenergy_rshock}), which starts to significantly deviate around the same time the explosion energy does, remaining then systematically higher for models \quadA{} and \quadB{} than for their axisymmetric counterparts.
This scenario is consistent with the results presented in \cite{muller2015}, which show that the more efficient growth of the Kelvin-Helmholtz instability in a 3D framework can lead to a higher efficiency in the neutrino-heating mechanism and hence more energetic explosions.
The faster expansion of the shock and the more energetic explosions found in model \aligneddip{} with respect to the case of an aligned quadrupolar magnetic field confirms the results presented in \cite{bugli2020}, which show that magnetic configurations of progressively higher multipolar order lead to weaker explosions and slower expanding ejecta.

The morphology of the expanding ejecta varies significantly depending on the initial configuration of the magnetic field (see \refig{fig:3d_entropy}).
While an aligned dipolar field leads to the formation of two symmetric and very well collimated outflows propagating through the stellar progenitor, in the case of an aligned quadrupolar field the ejecta are less collimated and exhibit a more complex structure, especially in the equatorial region.
A column of high entropy material can be found close to the rotational axis, with model \aligneddip{} having such material initially confined in a small region right behind the expanding shock front.
The unmagnetised model still produces prolate structures along the rotation axis, but the overall morphology of the ejecta is closer to a spherical distribution than to a polar outflow.
The outflows produced in model \tilteddip{} show an intermediate case: while the shock front is rather spherical, the entropy distribution close to the axis is very much collimated, similarly to what happens in model \quadA{}.

We can quantify the differences in ejecta morphology between different 3D models by decomposing the shock surface in its spherical harmonic components:
\begin{equation}\label{eq:shock_harm}
    \tilde{R}_{lm}=\int R_\mathrm{shock}(\theta,\phi) Y_{lm} \mathrm{d}\Omega,
\end{equation}
where $R_\mathrm{shock}$ is the shock radius and $Y_{lm}$ are the real spherical harmonics defined as
\begin{equation}
    Y_{lm}= \begin{cases}
                \sqrt{2}(-1)^m\mathrm{Im}[Y_l^{|m|}] &\quad\mathrm{if\ }m<0 \\
                Y_l^m &\quad\mathrm{if\ }m=0 \\
                \sqrt{2}(-1)^m\mathrm{Re}[Y_l^m] &\quad\mathrm{if\ }m>0,
            \end{cases}
\end{equation}
while
\begin{equation}
    Y_l^m=\sqrt{\frac{2l+1}{4\pi}\frac{(l-m)!}{(l+m)!}}P_l^m(\cos\theta)e^{im\phi},
\end{equation}
are the complex spherical harmonics, with $P_l^m$ the Legendre polynomial of order $(l,m)$.

The bottom right panel of \refig{fig:expenergy_rshock} presents the evolution of the $l=2$ axial harmonic component of the shock radius, confirming the fact that model \aligneddip{} produced the most collimated outflows, followed by the runs with aligned quadrupoles and then model \tilteddip{}, respectively.
Moreover, it shows that magnetised models producing fast shock expansions (i.e. those with aligned magnetic field) reach a peak in the degree of collimation around 200 ms p.b., relaxing then towards a lower and almost constant value.
Finally, axisymmetric models display significantly more prolate shocks than their 3D counterparts.
In the case of aligned quadrupolar fields, there is a fast rise of $\tilde{R}_{20}$ for the 2D simulations coinciding with the onset of the shock's expansion, whereas for models \quadA{} and \quadB{} it remains roughly constant.
For the hydrodynamic case, the 3D version is characterised by a flattening of the shock surface between 100 and 400 ms p.b., after which it becomes rather spherical.
On the other hand, the shock surface of model \hydrotwod{} is never oblate, and evolves into a more prolate shape when the onset of the explosion occurs.
These findings point to the fact that the assumption of axisymmetry tends to overestimate the degree of collimation of the outflows launched during the explosion, as a result of preventing the dynamics of the system from developing non-axisymmetric structures.

%%%%%%%%%%%%%%%%%%%%%%%%%%%
%%%%%%%%%%%%%%%%%%%%%%%%%%%
%%%%%%%%%%%%%%%%%%%%%%%%%%%
%%%%%%%%%%%%%%%%%%%%%%%%%%%
%%%%%%%%%%%%%%%%%%%%%%%%%%%
%%%%%%%%%%%%%%%%%%%%%%%%%%%
%%%%%%%%%%%%%%%%%%%%%%%%%%%
%%%%%%%%%%%%%%%%%%%%%%%%%%%
%%%%%%%%%%%%%%%%%%%%%%%%%%%
%%%%%%%%%%%%%%%%%%%%%%%%%%%
%%%%%%%%%%%%%%%%%%%%%%%%%%%
%%%%%%%%%%%%%%%%%%%%%%%%%%%
%%%%%%%%%%%%%%%%%%%%%%%%%%%
%%%%%%%%%%%%%%%%%%%%%%%%%%%
%%%%%%%%%%%%%%%%%%%%%%%%%%%
%%%%%%%%%%%%%%%%%%%%%%%%%%%
%%%%%%%%%%%%%%%%%%%%%%%%%%%
%%%%%%%%%%%%%%%%%%%%%%%%%%%

\subsection{PNS evolution}

\begin{figure}
    \centering
    \includegraphics[width=0.45\textwidth]{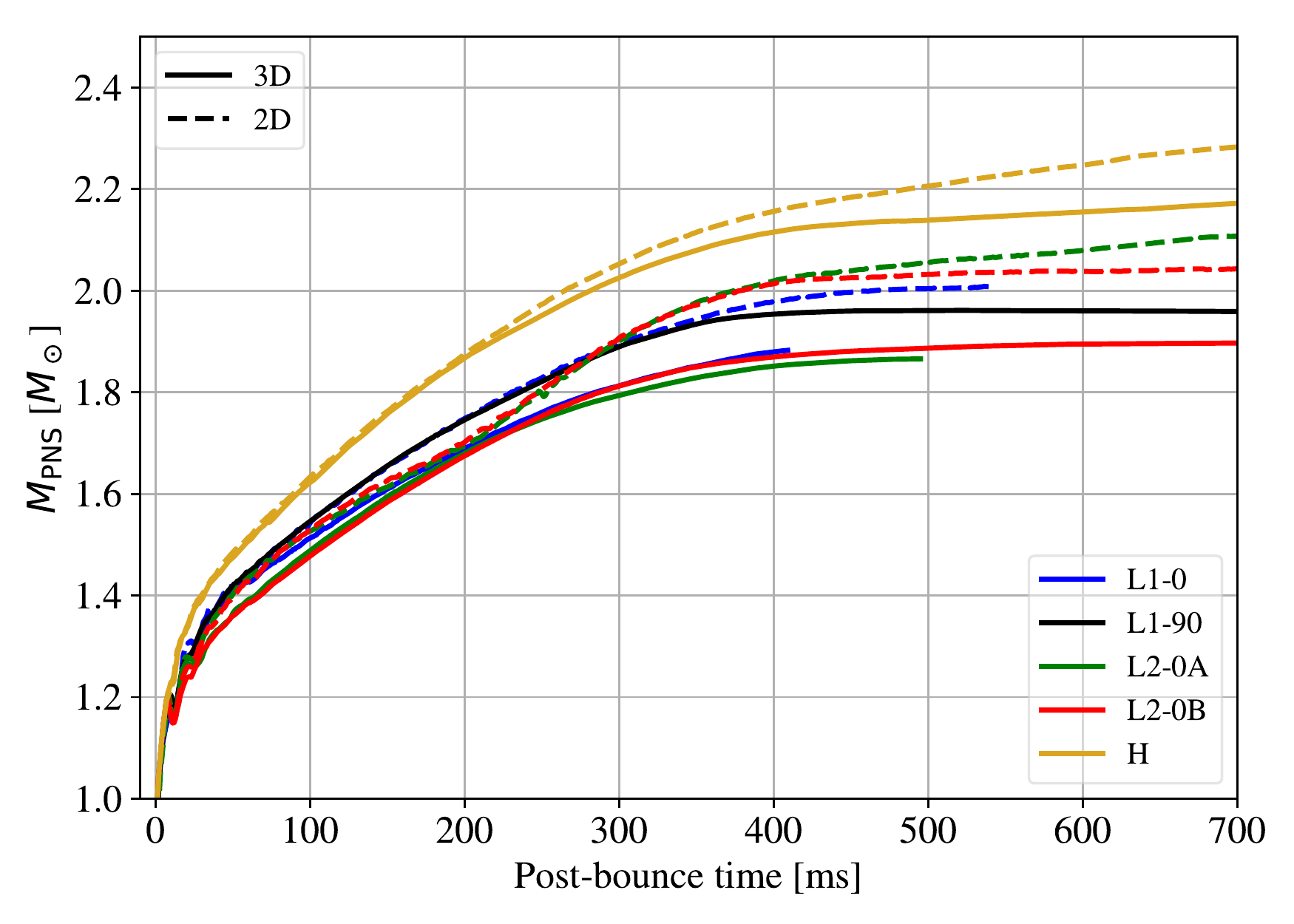}
    \includegraphics[width=0.45\textwidth]{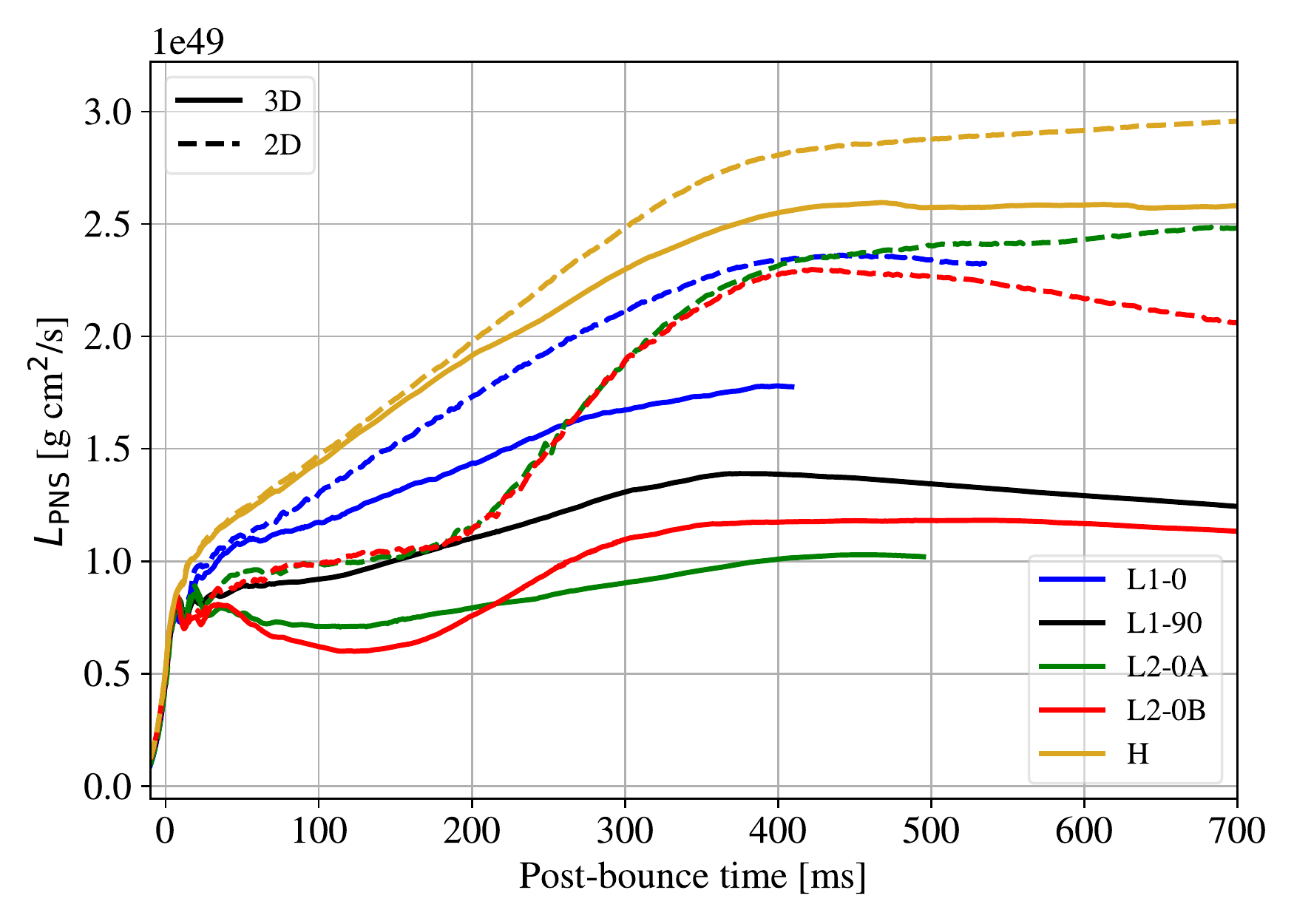}
    \caption{PNS mass (top panel) and total angular momentum (bottom) over time.}
    \label{fig:PNS}
\end{figure}

\begin{figure}
    \centering
    \includegraphics[width=0.45\textwidth]{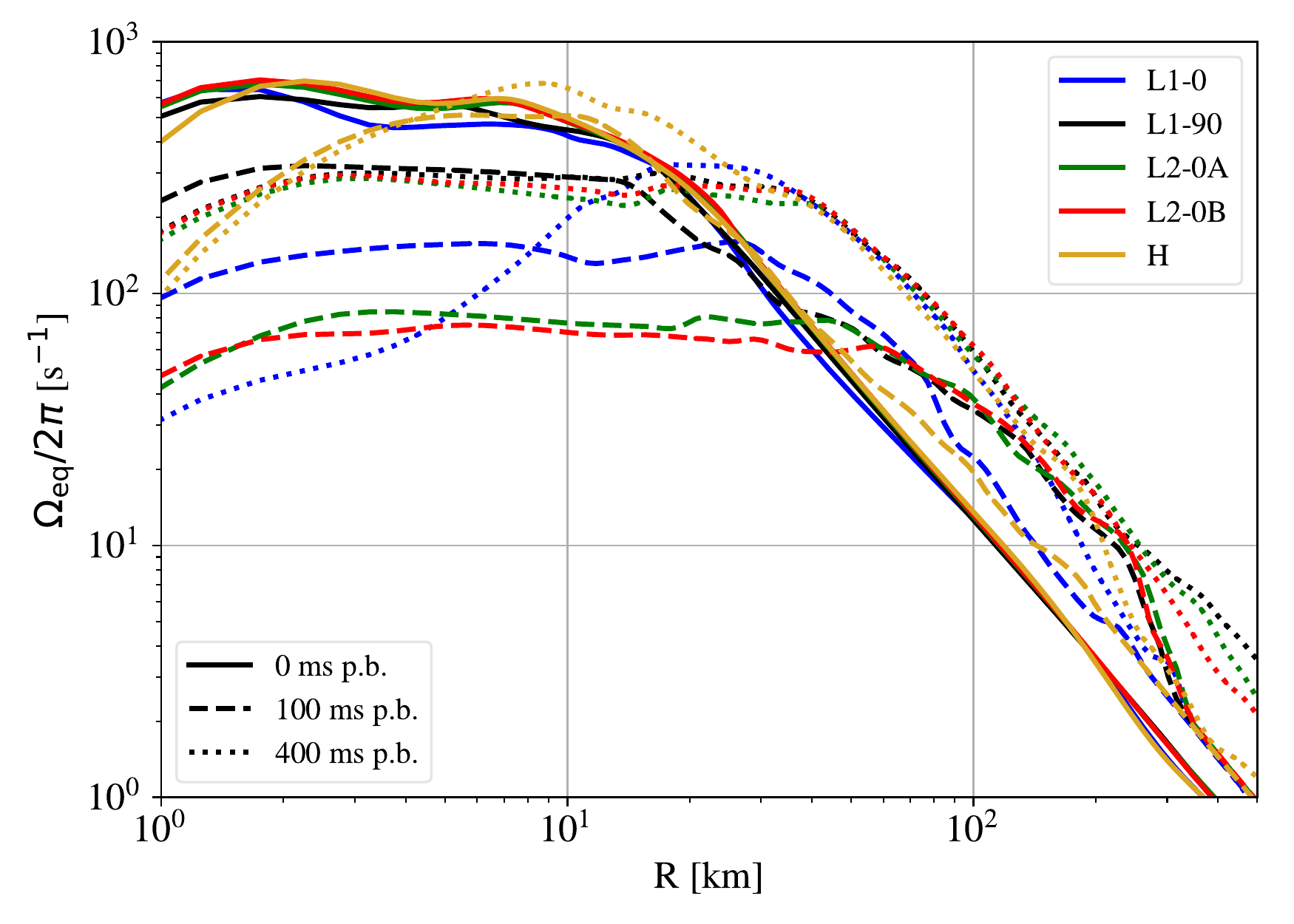}
    \includegraphics[width=0.45\textwidth]{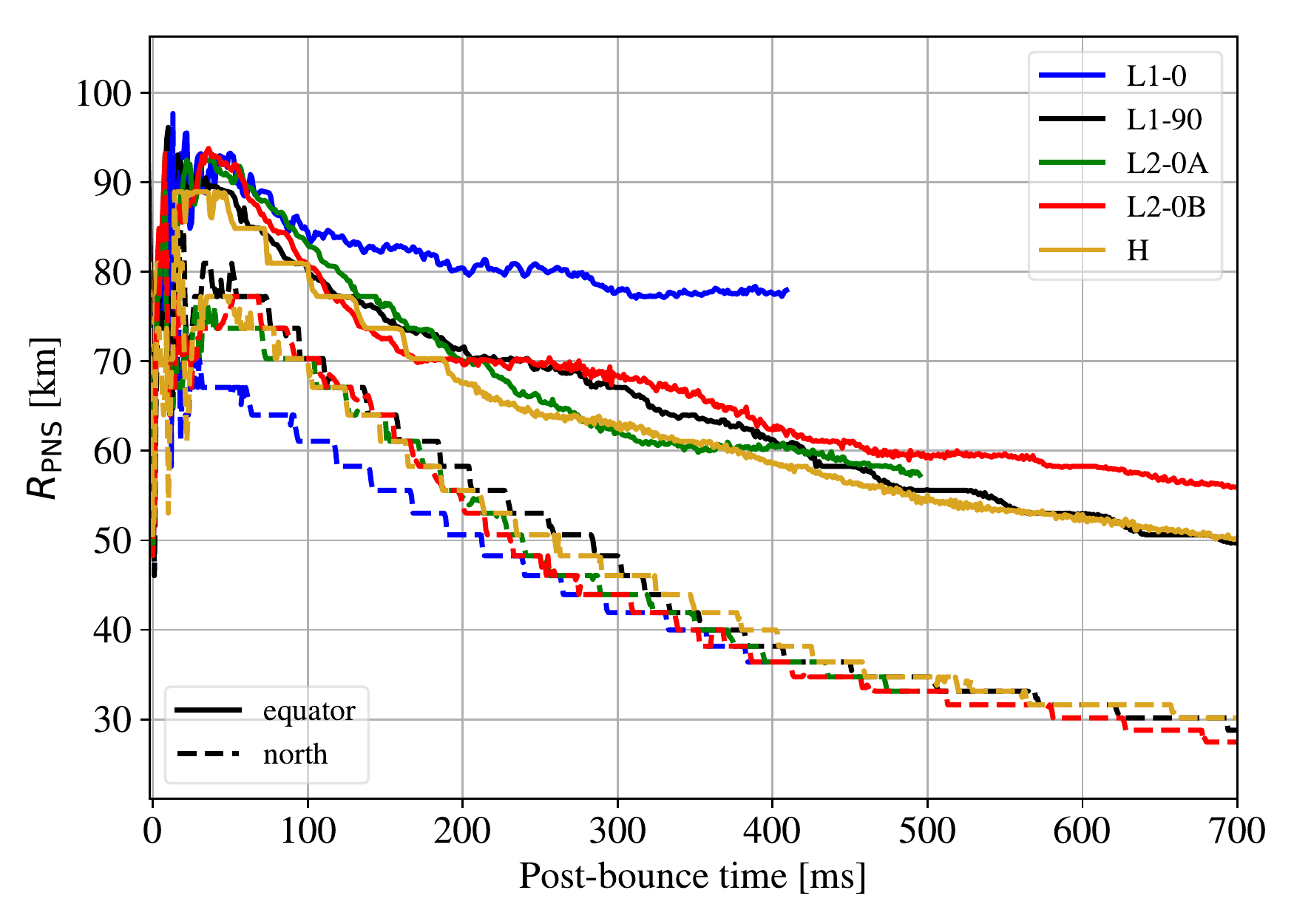}
    \caption{Equatorial radial profile of the angular velocity (top panel) and time series of the azimuthally averaged PNS radii (bottom).}
    \label{fig:omega_r}
\end{figure}

\begin{figure}
    \centering
    \includegraphics[trim=0.2cm 2.7cm 0 0, clip, width=0.45\textwidth]{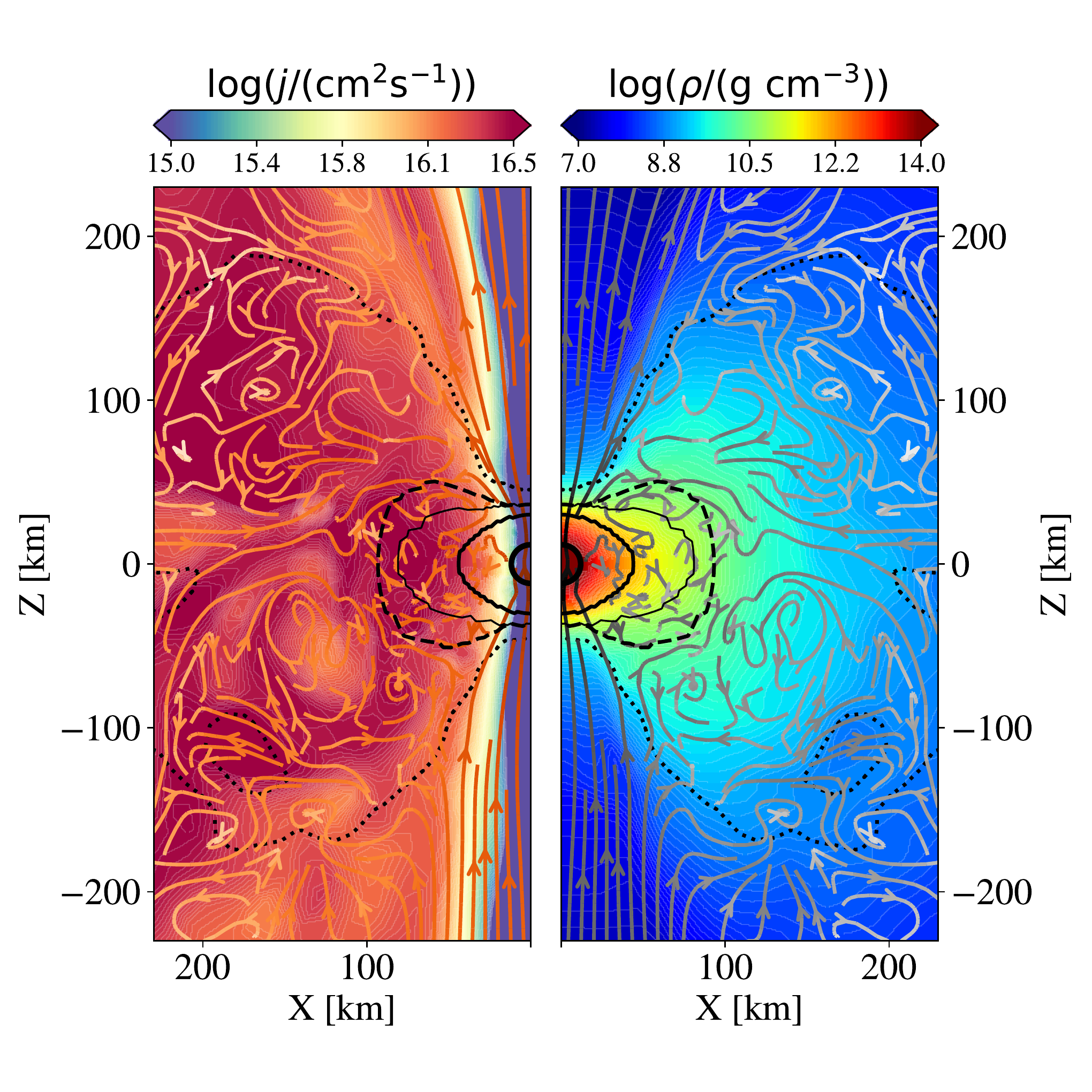}
    \includegraphics[trim=0 0 0 3.5cm, clip, width=0.45\textwidth]{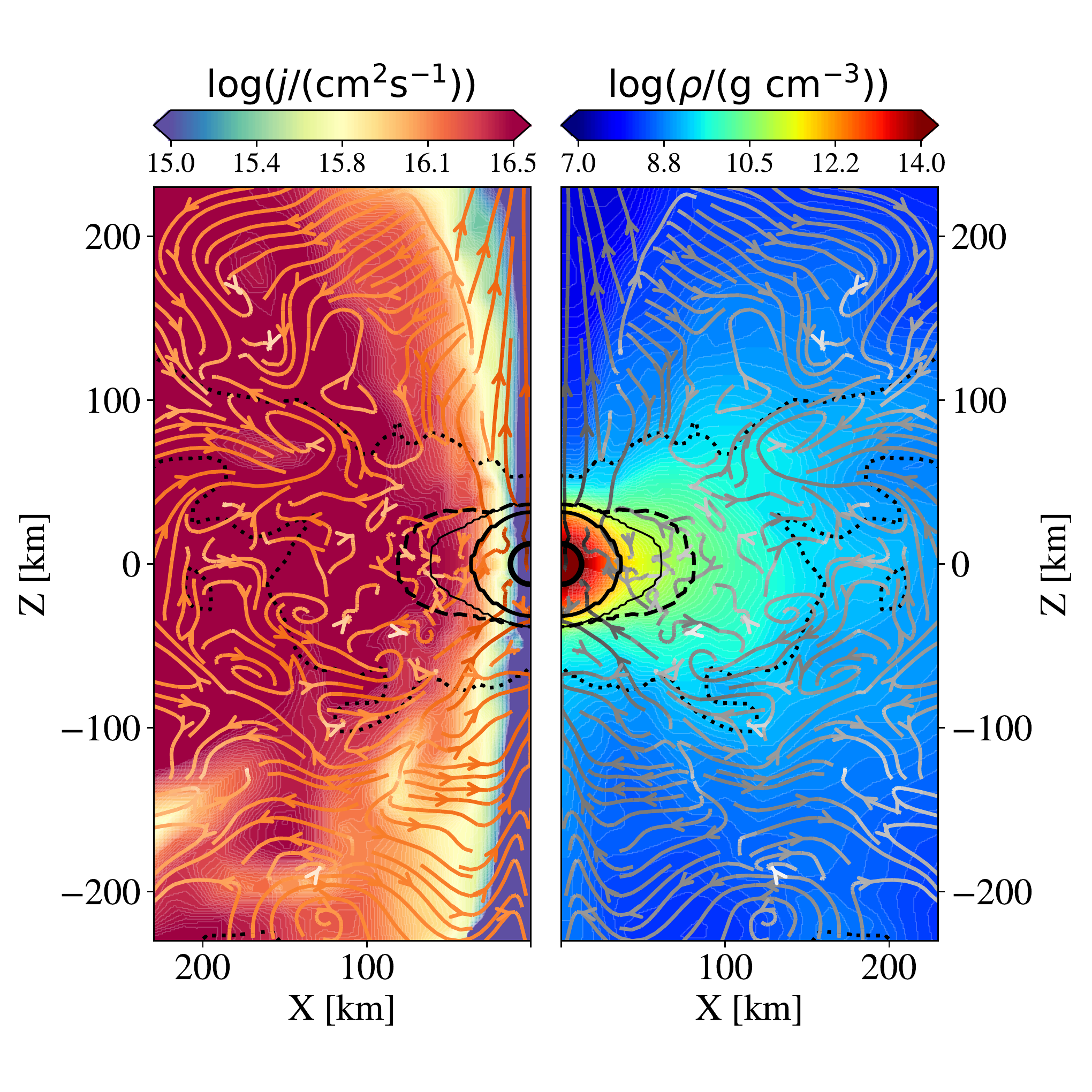}
    \caption{Meridional cuts of specific angular momentum (left) and mass density (right) for model \aligneddip{} (top) and \quadA{} (bottom) at $t=400$ ms p.b. The dotted lines delimit the gravitationally bound ejecta, the dashed lines are the neutrino-spheres for the electron neutrinos and the three solid lines are contours of constant density $10^{11}$, $10^{12}$ and $10^{14}$ g/cm$^3$.}
    \label{fig:j-rho}
\end{figure}

We now focus on the evolution of the PNS, whose volume is defined as the region of the domain where the matter density exceeds the threshold value of $10^{11}$ g/cm$^3$.
Among all models presented in this work, the hydrodynamic one produces the most massive PNS, reaching a mass of $\sim2.2M_\odot$ around 700 ms p.b. (top panel of \refig{fig:PNS}).
On the other hand, while configurations with aligned magnetic fields produce lighter PNS (with the quadrupolar case plateauing close to $1.9M_\odot$), model \tilteddip{} produces an intermediate scenario, with $M_\mathrm{PNS}$ almost reaching $2M_\odot$ at 400 ms p.b. and remaining constant for the rest of the simulation.
The correspondent axisymmetric models (dashed lines) show systematically more massive PNSs, with a clear deviation from the three-dimensional case occurring around 200 ms p.b. 
This is again reminiscent of the non-magnetised results presented in \cite{muller2015}, which show that the more efficient growth of the Kelvin-Helmholtz instability in 3D prevents continued accretion onto the PNS.

If we look at the total angular momentum contained in the forming PNS $L_\mathrm{PNS}$ (bottom panel in \refig{fig:PNS}) we can see a similar dependence on the initial magnetic field, since a less massive PNS tends to also rotate more slowly.
However, the magnetised models show significantly different evolution of the PNS angular momentum, with models \quadA{} and \quadB{} presenting slower rotation than the case with an aligned dipolar field and model \tilteddip{} being an intermediate case.
The more efficient extraction of angular momentum by a quadrupolar field or an inclined dipole compared to the aligned dipole can be understood by the non-vanishing radial magnetic field in the equatorial plane (see below the discussion of rotation profiles).
All the correspondent axisymmetric models produce a faster rotating PNS (in the case of model \quadA{} more than a factor 2) and deviate from the three-dimensional case after a few tens of ms from the shock formation.
Note that the change of slope occurring around 400 ms p.b. is connected to the accretion of the iron core surface, beyond which there is a steep decrease of the specific angular momentum.

The radial profile of the angular velocity at the equator provides some interesting insights on the spin-down of the PNS and the transport of angular momentum (\refig{fig:omega_r}).
At 100 ms p.b. (dashed lines in the top panel) these profiles are already rather different among the various models, with the quadrupolar ones displaying more extended and slower inner cores in solid-body rotation, with a transition to a decreasing profile that between 50 and 250 km has a shallower slope compared to models \hydro{} and \aligneddip{}.
Models with dipolar fields have faster and smaller rigidly rotating inner region (to a larger extent for model \tilteddip{}), while the simulation without magnetic fields does not show any region with a flat profile of $\Omega(r)$.
Such profiles suggest that the transport of angular momentum in the equatorial plane of the PNS is much more efficient with a quadrupolar field, rather than a dipolar one.
Later after bounce (400 ms, dotted lines) the situation appears to have further evolved.
While models with a quadrupolar field or a tilted dipole have an inner core spinning at a frequency of $\sim 300$ Hz and extending up to 40 km, the PNS in model \aligneddip{} has a strongly spun-down central core.
However, beyond 100 km from the centre the model with an aligned dipolar field shows a steeper decrease of the rotation profile, pointing once again to a less efficient transport of angular momentum in the radial direction from the PNS to its immediate surroundings on the equatorial plane.
Finally, model \hydro{} shows only a general increase with time of the angular velocity at radii larger than $\sim5$ km, which is consistent with the advection of angular momentum by the accreting material and the absence of outward transport mediated by magnetic fields.

\begin{figure*}
    % \centering
    \includegraphics[width=0.8\textwidth]{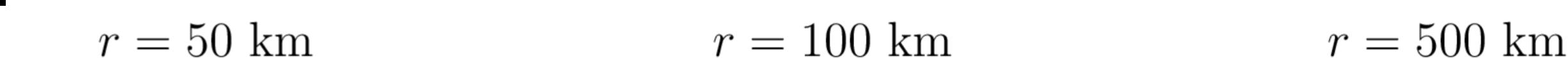}\\
    \includegraphics[width=0.04\textwidth]{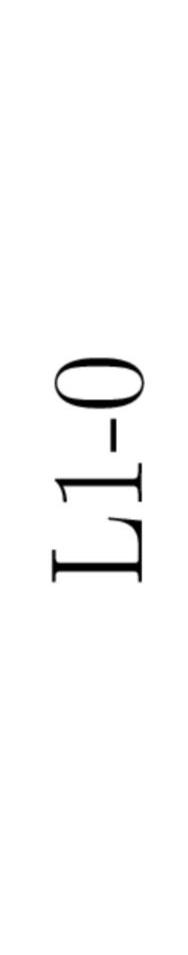}
    \includegraphics[width=0.31\textwidth]{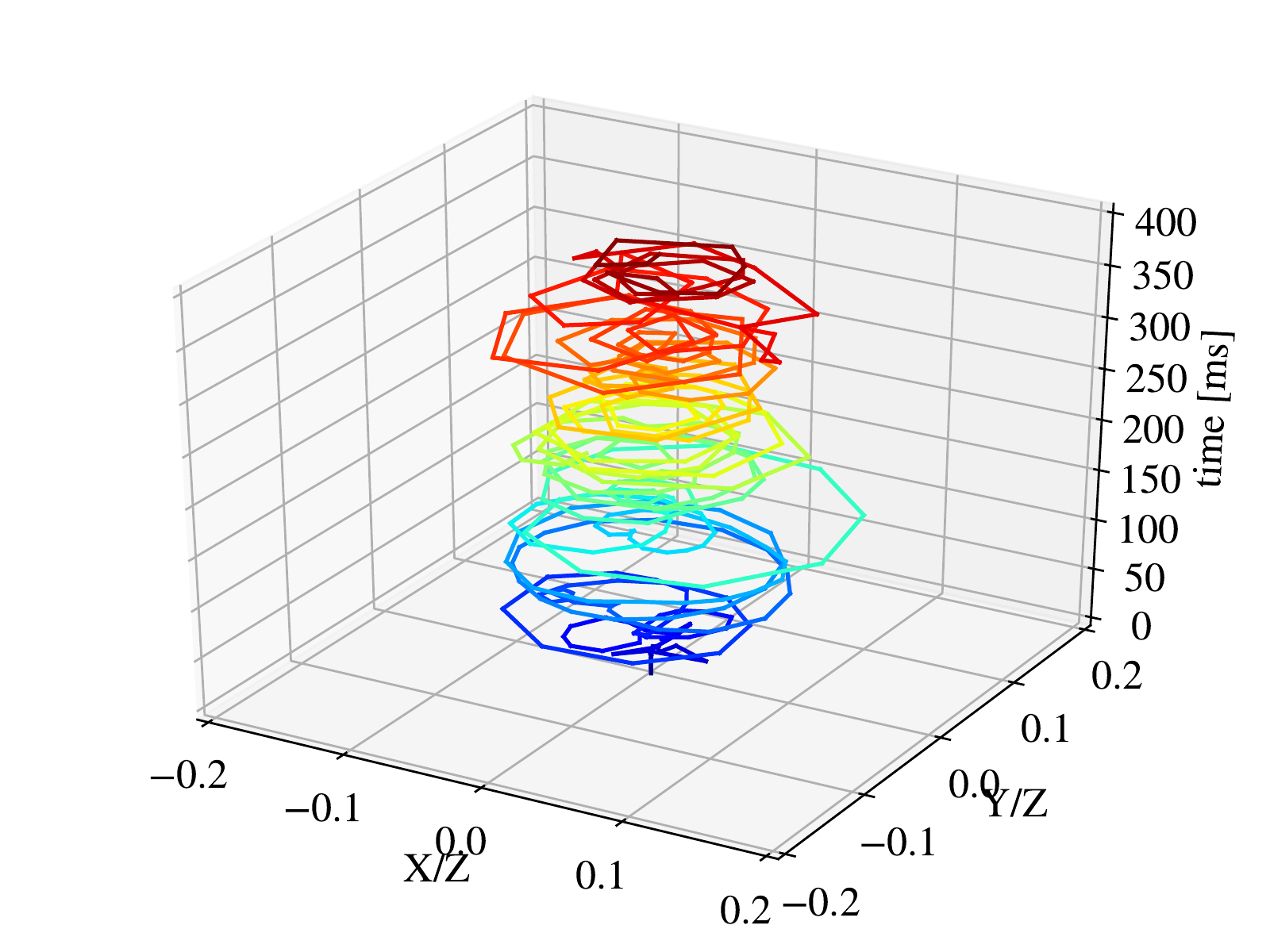}
    \includegraphics[width=0.31\textwidth]{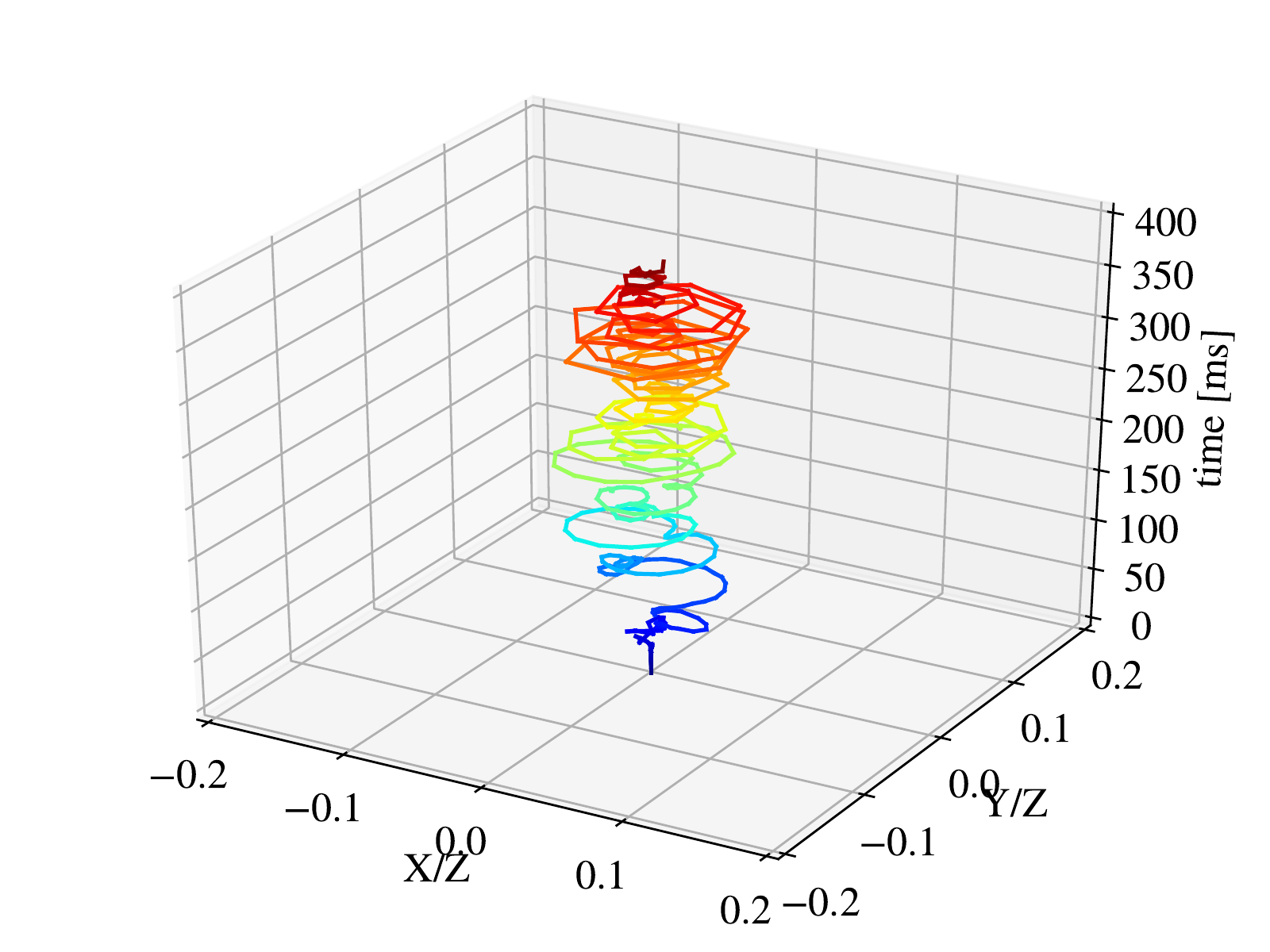}
    \includegraphics[width=0.31\textwidth]{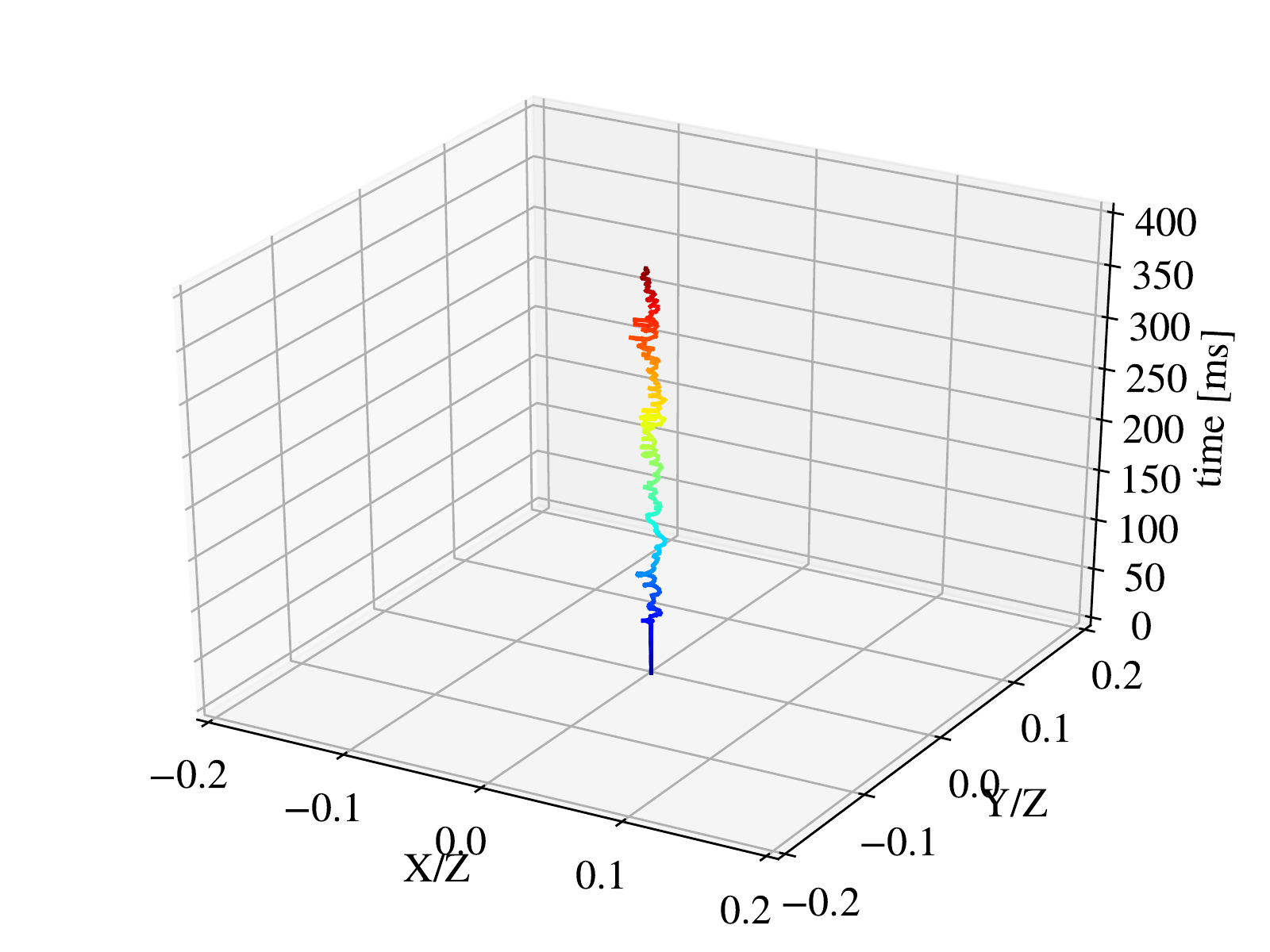}
    \includegraphics[width=0.04\textwidth]{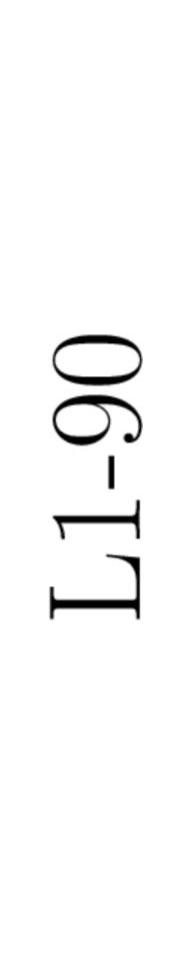}
    \includegraphics[width=0.31\textwidth]{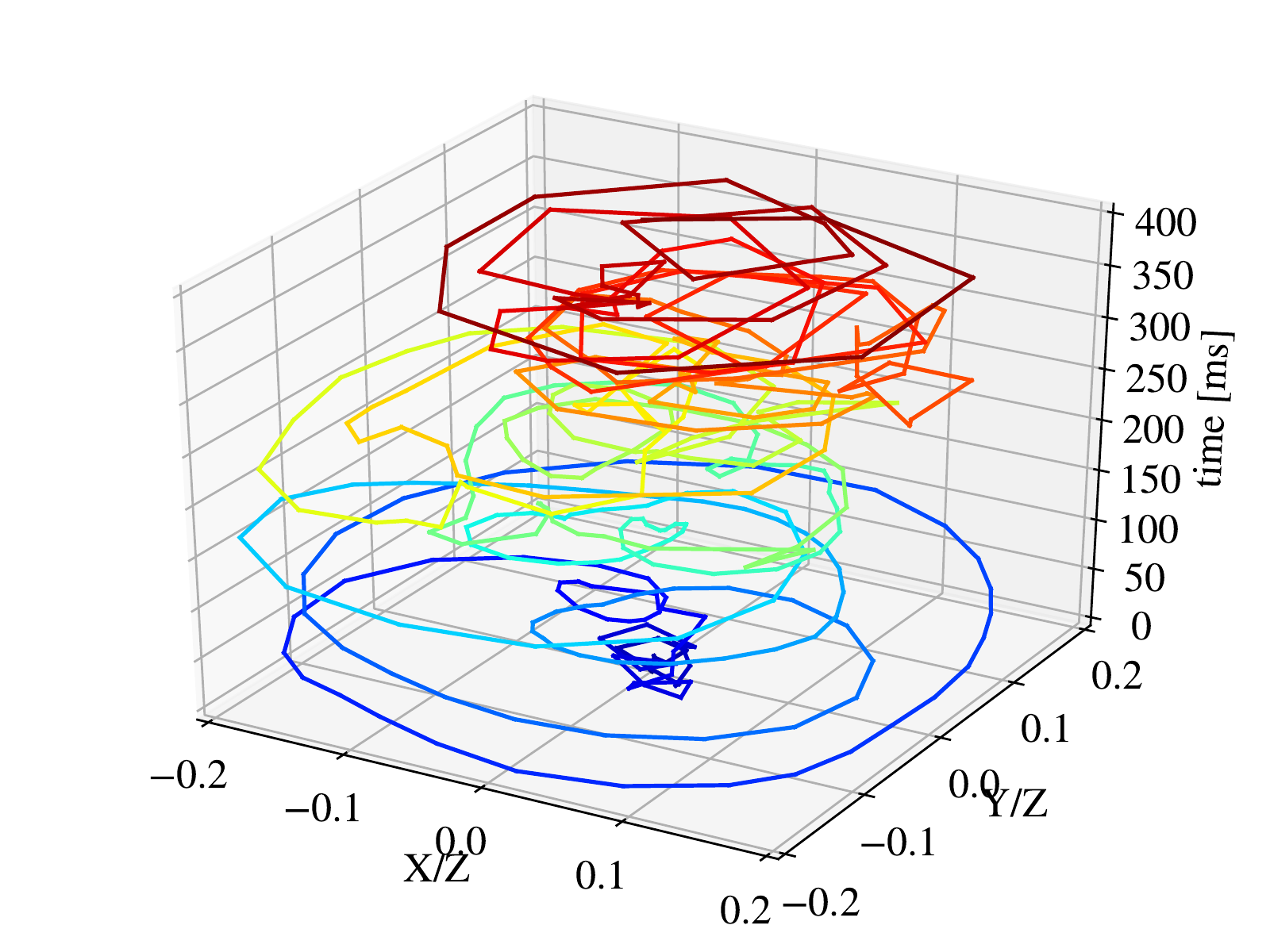}
    \includegraphics[width=0.31\textwidth]{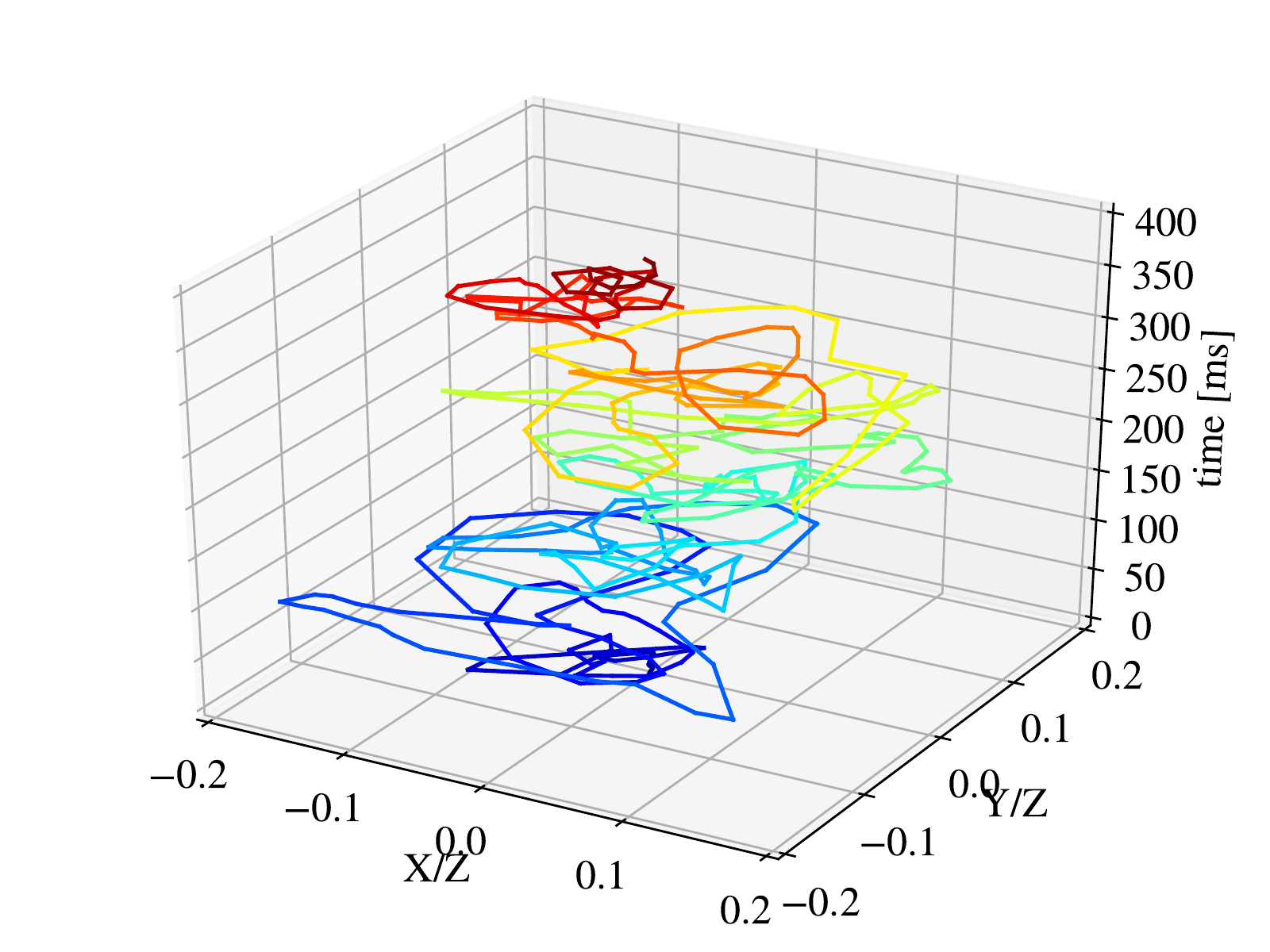}
    \includegraphics[width=0.31\textwidth]{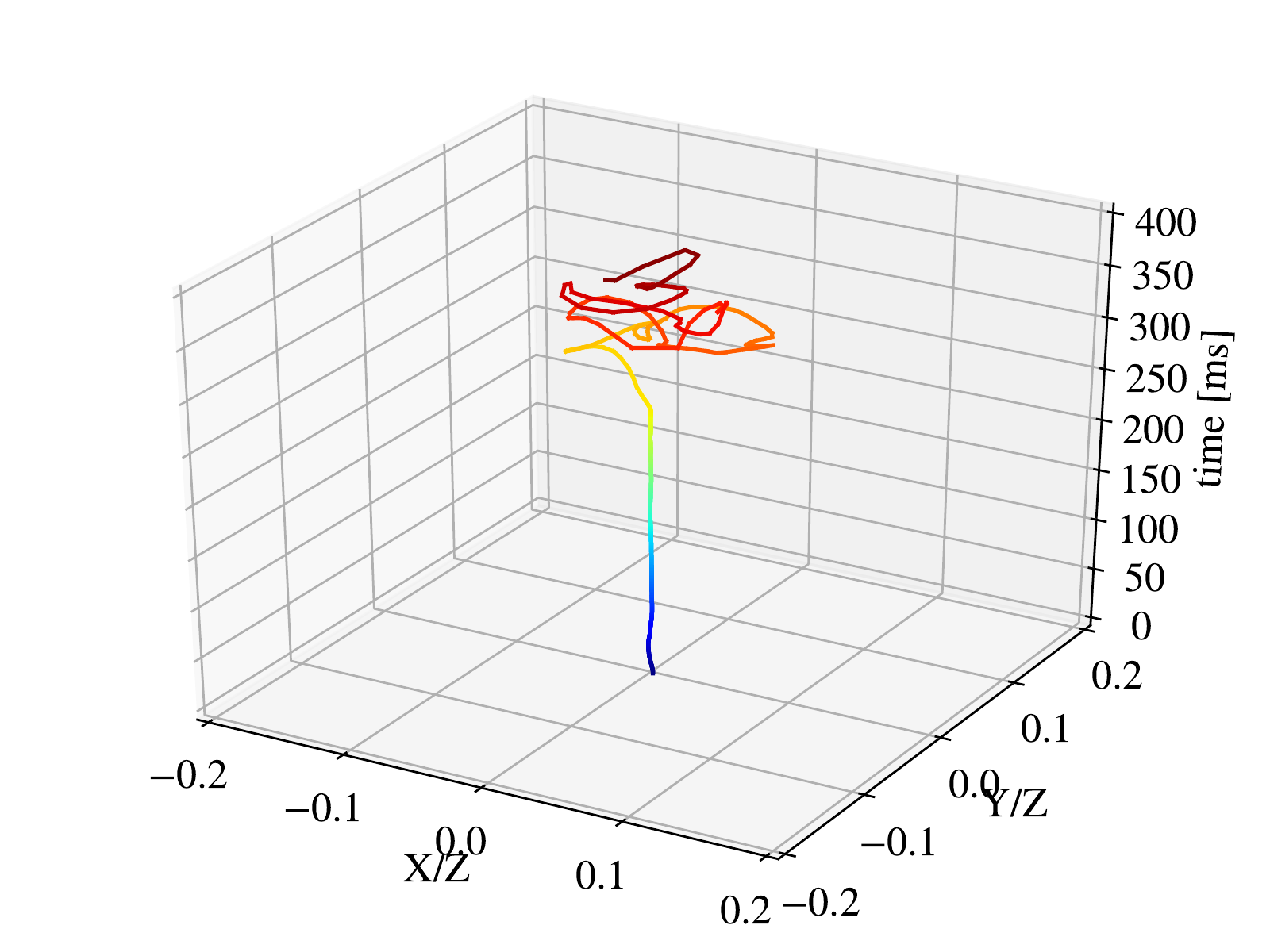}
    \includegraphics[width=0.04\textwidth]{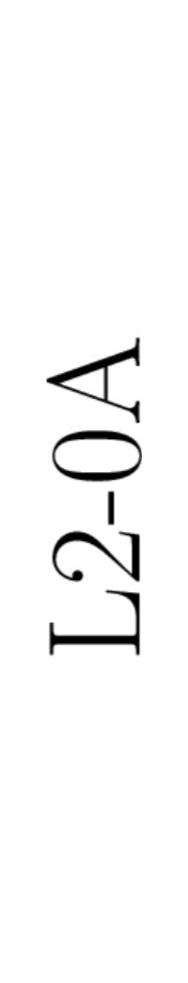}
    \includegraphics[width=0.31\textwidth]{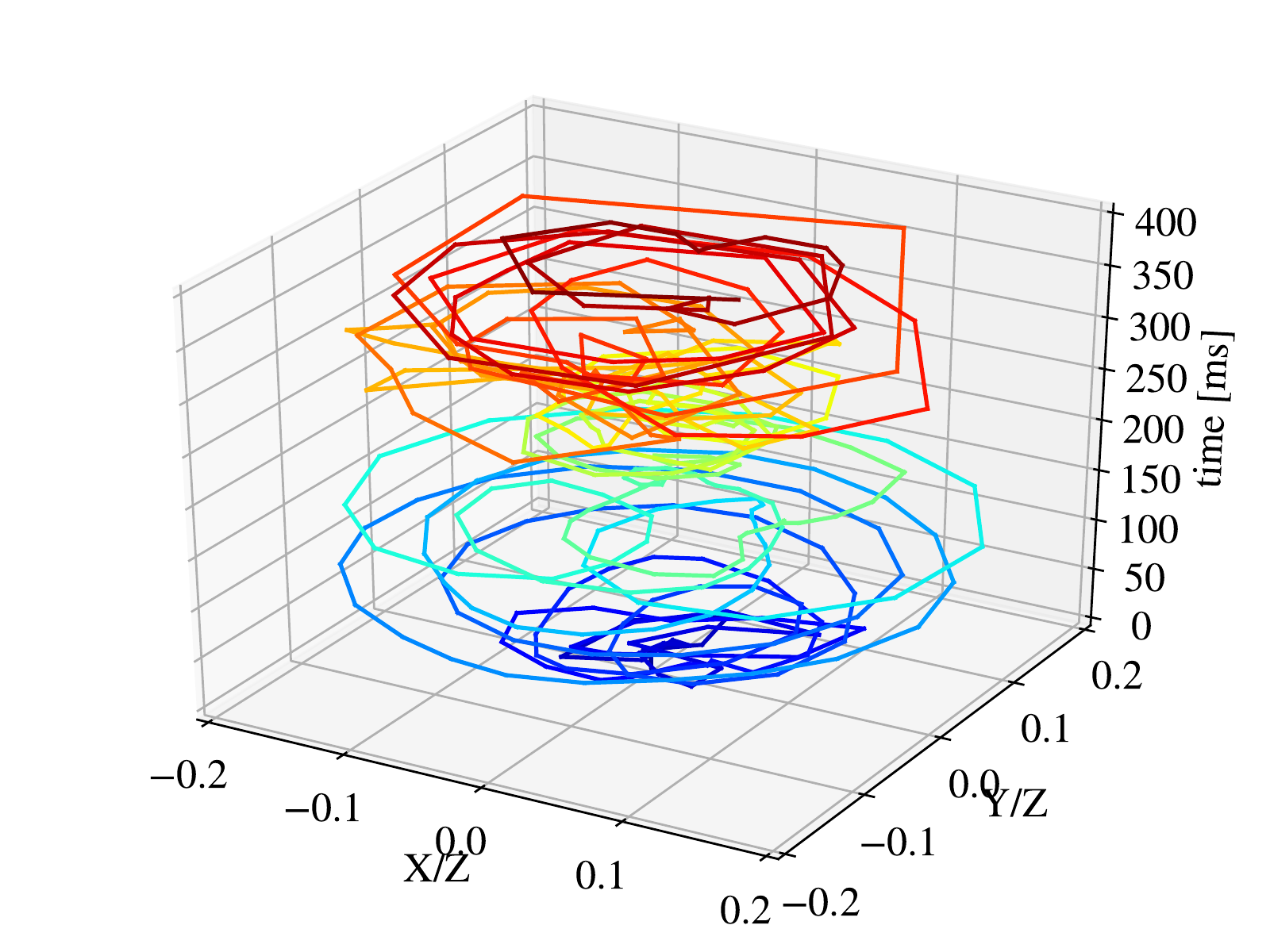}
    \includegraphics[width=0.31\textwidth]{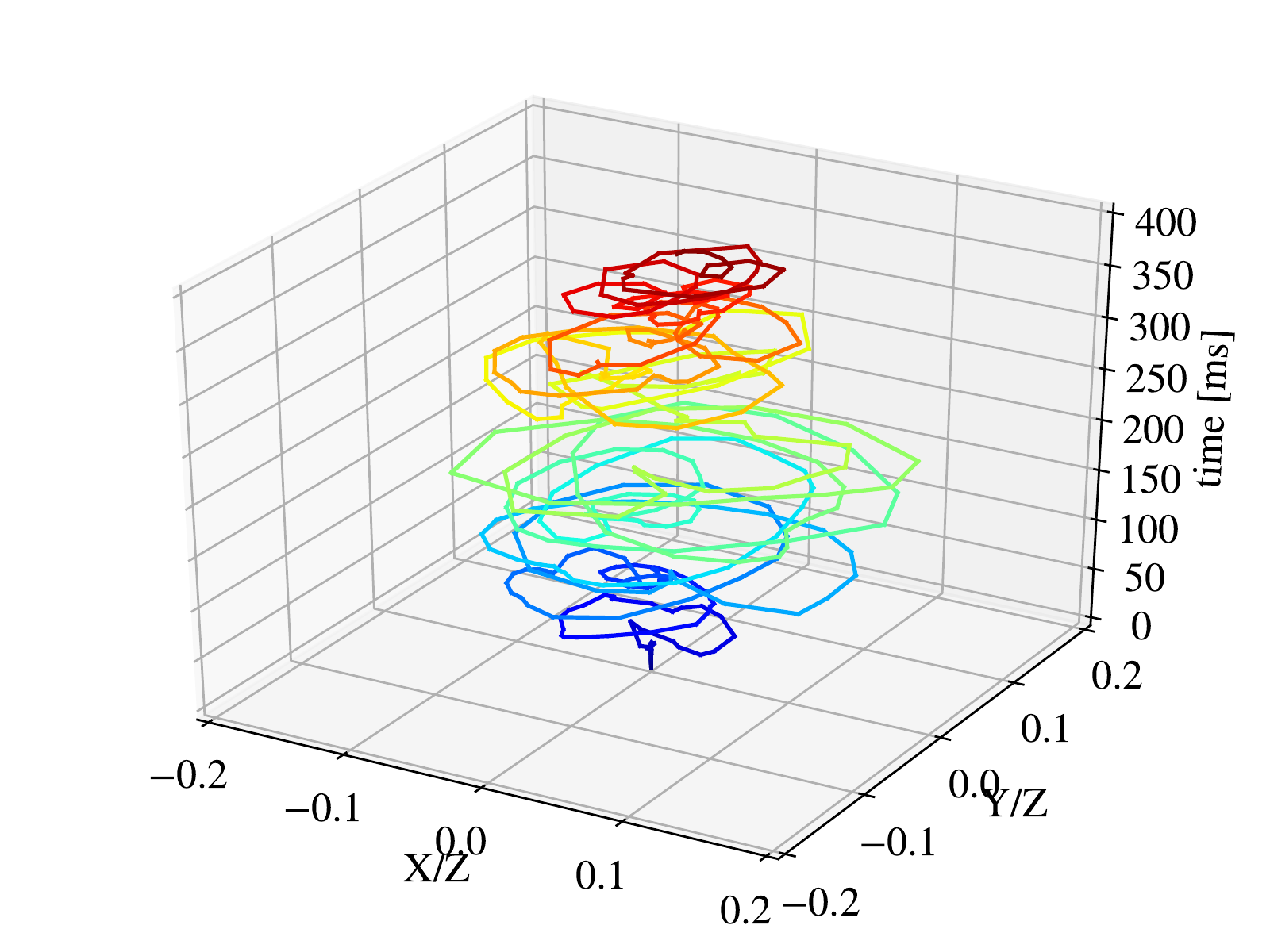}
    \includegraphics[width=0.31\textwidth]{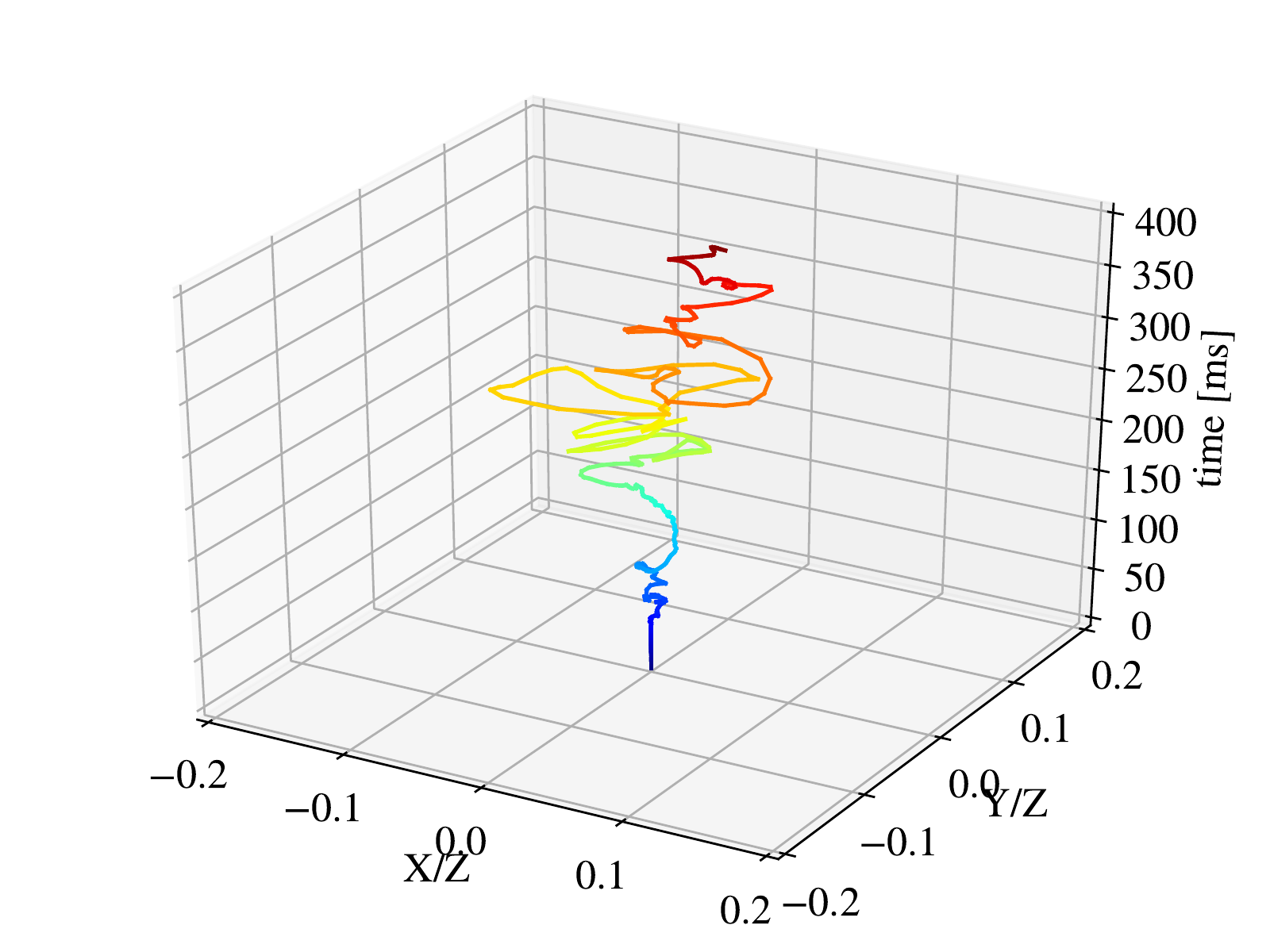}
    \caption{Space-time trajectories of the jet barycenter  in the x-y plane at a distance of $z=50$, 100 and 500 km (first to last columns, respectively) for models \aligneddip{}, \tilteddip{} and \quadA{} (first to last row).}
    \label{fig:jet_barycenter}
\end{figure*}

Despite having the most angular momentum among magnetised models, the PNS of run \aligneddip{} is not the fastest spinning.
This discrepancy is due to its different shape, as we can see from the bottom panel of \refig{fig:omega_r}.
The PNS of the model with an aligned dipolar field is, at any given time, the most oblate one, having the largest equatorial radius and the smallest polar one.
This leads to a larger moment of inertia, which compensates for the slower inner core and leads to a higher total angular momentum than model \tilteddip{}.
This is even more evident if we look at a meridional section of the PNS at 400 ms p.b. (\refig{fig:j-rho}).
The PNS surface (i.e. the region where density reaches the value of $10^{11}$ g/cm$^3$) is much more oblate in model \aligneddip{} than \quadA{}, with the neutrino-sphere also having a more distinctive peanut-shape.
On the other hand, the surrounding of the PNS appear to have a stronger rotational support in the case of the quadrupolar field, as the gravitationally bound region close to the PNS is quite smaller and surrounded by more rarefied material with higher specific angular momentum.

%%%%%%%%%%%%%%%%%%%%%%%%%%%
%%%%%%%%%%%%%%%%%%%%%%%%%%%
%%%%%%%%%%%%%%%%%%%%%%%%%%%
%%%%%%%%%%%%%%%%%%%%%%%%%%%
%%%%%%%%%%%%%%%%%%%%%%%%%%%
%%%%%%%%%%%%%%%%%%%%%%%%%%%
%%%%%%%%%%%%%%%%%%%%%%%%%%%
%%%%%%%%%%%%%%%%%%%%%%%%%%%
%%%%%%%%%%%%%%%%%%%%%%%%%%%
%%%%%%%%%%%%%%%%%%%%%%%%%%%
%%%%%%%%%%%%%%%%%%%%%%%%%%%
%%%%%%%%%%%%%%%%%%%%%%%%%%%
%%%%%%%%%%%%%%%%%%%%%%%%%%%
%%%%%%%%%%%%%%%%%%%%%%%%%%%
%%%%%%%%%%%%%%%%%%%%%%%%%%%
%%%%%%%%%%%%%%%%%%%%%%%%%%%
%%%%%%%%%%%%%%%%%%%%%%%%%%%
%%%%%%%%%%%%%%%%%%%%%%%%%%%

\subsection{Kink instability}

The stability of the magnetised outflows produced in MHD core-collapse numerical models has received the attention of many studies in the last few years.
\cite{mosta2014b} showed for the first time that the outflows produced in a fully three-dimensional magnetorotational explosion are prone to develop the so-called \emph{kink instability} \citep[e.g.,][]{Eichler1993,Begelman1998}, i.e. a large-scale non-axisymmetric instability that can disrupt the coherence of the polar jets and possibly prevent their propagation through the stellar progenitor.
While the recent work in \cite{kuroda2020} corroborates this scenario, other three-dimensional studies managed to produce successful MHD driven explosions with bipolar outflows retaining their coherent structure up to one second p.b. and over thousands of kilometers \citep{obergaulinger2020a,obergaulinger2021a}.

While it is clear from \refig{fig:expenergy_rshock} and \refig{fig:3d_entropy} that all our models produce successful explosions and magnetised outflows, it is important to quantitatively estimate the growth of the kink instability in the region close to the axis and make a comparison with the existing literature.
The displacement of the jet is tracked by the coordinates of its barycenter, defined as \citep{mosta2014b}:
\begin{equation}
    x_c^i=\frac{\int x^i P_\mathrm{mag} \mathrm{~d}S}{\int P_\mathrm{mag} \mathrm{~d}S},
\end{equation}
where $P_\mathrm{mag}=B^2/2$ is the magnetic pressure, $x^i$ stands for the Cartesian coordinates $\{x,y,z\}$ and the integrals extend to an horizontal circular surface $S$ of radius $r_S=\mathrm{max}(|z|,50\mathrm{ km})$ centred at the axis.

\refig{fig:jet_barycenter} shows the space-time trajectories of the jet's barycenter in the x-y plane at different distances $z=50$, 100 and 500 km from the center.
Model \aligneddip{} shows a remarkable regularity in the circular motion of the jet, whose displacement with respect to the rotation axis never exceeds 10\% of its vertical coordinate $z$.
Moreover, the motions due to the development of the kink instability become less important the further we get from the PNS.
The models with quadrupolar fields and an equatorial dipole, on the other hand, produce oscillations of significantly higher amplitude, although they tend in these cases as well to become less relevant at longer distances from the PNS.
Model \tilteddip{} seem to produce the least regular trajectories (especially at 100 km form the center), which could be due to the inherent non-axisymmetric structure of its initial magnetic field.

\begin{figure}
    \centering
    \includegraphics[width=0.45\textwidth]{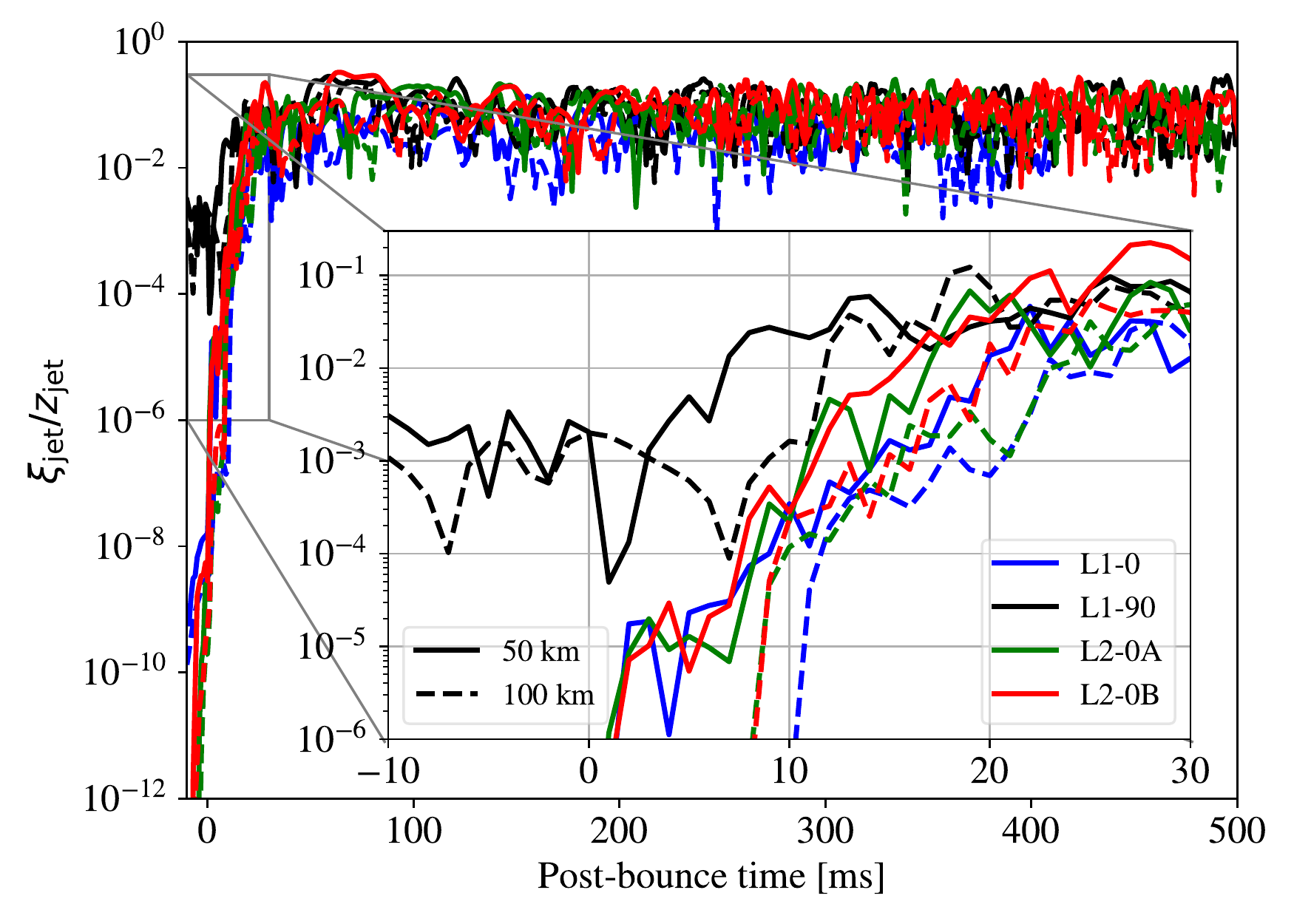}
    \caption{Jet barycenter displacement over time at a distance of $z=50$ km (solid curves) and 100 km (dashed curves).}
    \label{fig:jet_displacement}
\end{figure}

As we can see from \refig{fig:jet_displacement}, in the first few tens of millisecond there is an exponential growth of the displacement of the jet's barycenter in every magnetised explosion.
Model \tilteddip{} shows high values of $\xi$ even before bounce, as it is inherently non-axisymmetric.
It is interesting to note, however, that even in this case the displacement of the jet's barycenter saturates around values that do not substantially differ from those produced in other magnetised models.
The case of an aligned dipole and quadrupoles show instead a clear linear phase, having initial perturbations at bounce with amplitude in between $10^{-9}$ and $10^{-8}$.
For these 3 models the displacement at 50 km from the center (solid curves) increases at a similar rate with an $e$-folding time of approximately 2.2 ms, reaching saturation after about 30 ms p.b.
Such a growth rate is of the same order of magnitude as the one found in \cite{mosta2014a}, although its exact value depends on the intensity of the toroidal magnetic field, which in return is directly affected by the rotation profile employed and the strength of the initial magnetic field.
Our results on the kink instability are in agreement with the models presented in the recent literature \citep{mosta2014a,kuroda2020,obergaulinger2021a} on the fact that 3D simulations are prone to develop non-axisymmetric modes on very short time-scales.
However, in our simulations such a phenomenon does not prevent the magnetorotational mechanism from launching an explosion in the form of collimated outflows, in agreement with \citet{obergaulinger2021a} but in contrast to \citet{mosta2014a,kuroda2020}.
The cause for this different behaviour could lie in many different aspects of our simulations: progenitor, magnetic field strength and structure, numerical grid, etc.
Further work is required to pinpoint the main ingredients determining whether an explosion with collimated outflows can be launched.
In particular, a comparison study between different codes employing the same initial conditions could shed some light on the impact of numerical aspects such as topology of the grid on the development of the kink instability in core-collapse supernovae.

%%%%%%%%%%%%%%%%%%%%%%%%%%%
%%%%%%%%%%%%%%%%%%%%%%%%%%%
%%%%%%%%%%%%%%%%%%%%%%%%%%%
%%%%%%%%%%%%%%%%%%%%%%%%%%%
%%%%%%%%%%%%%%%%%%%%%%%%%%%
%%%%%%%%%%%%%%%%%%%%%%%%%%%
%%%%%%%%%%%%%%%%%%%%%%%%%%%
%%%%%%%%%%%%%%%%%%%%%%%%%%%
%%%%%%%%%%%%%%%%%%%%%%%%%%%
%%%%%%%%%%%%%%%%%%%%%%%%%%%
%%%%%%%%%%%%%%%%%%%%%%%%%%%
%%%%%%%%%%%%%%%%%%%%%%%%%%%
%%%%%%%%%%%%%%%%%%%%%%%%%%%
%%%%%%%%%%%%%%%%%%%%%%%%%%%
%%%%%%%%%%%%%%%%%%%%%%%%%%%
%%%%%%%%%%%%%%%%%%%%%%%%%%%
%%%%%%%%%%%%%%%%%%%%%%%%%%%
%%%%%%%%%%%%%%%%%%%%%%%%%%%

\begin{figure}
    \centering
    \includegraphics[width=0.45\textwidth]{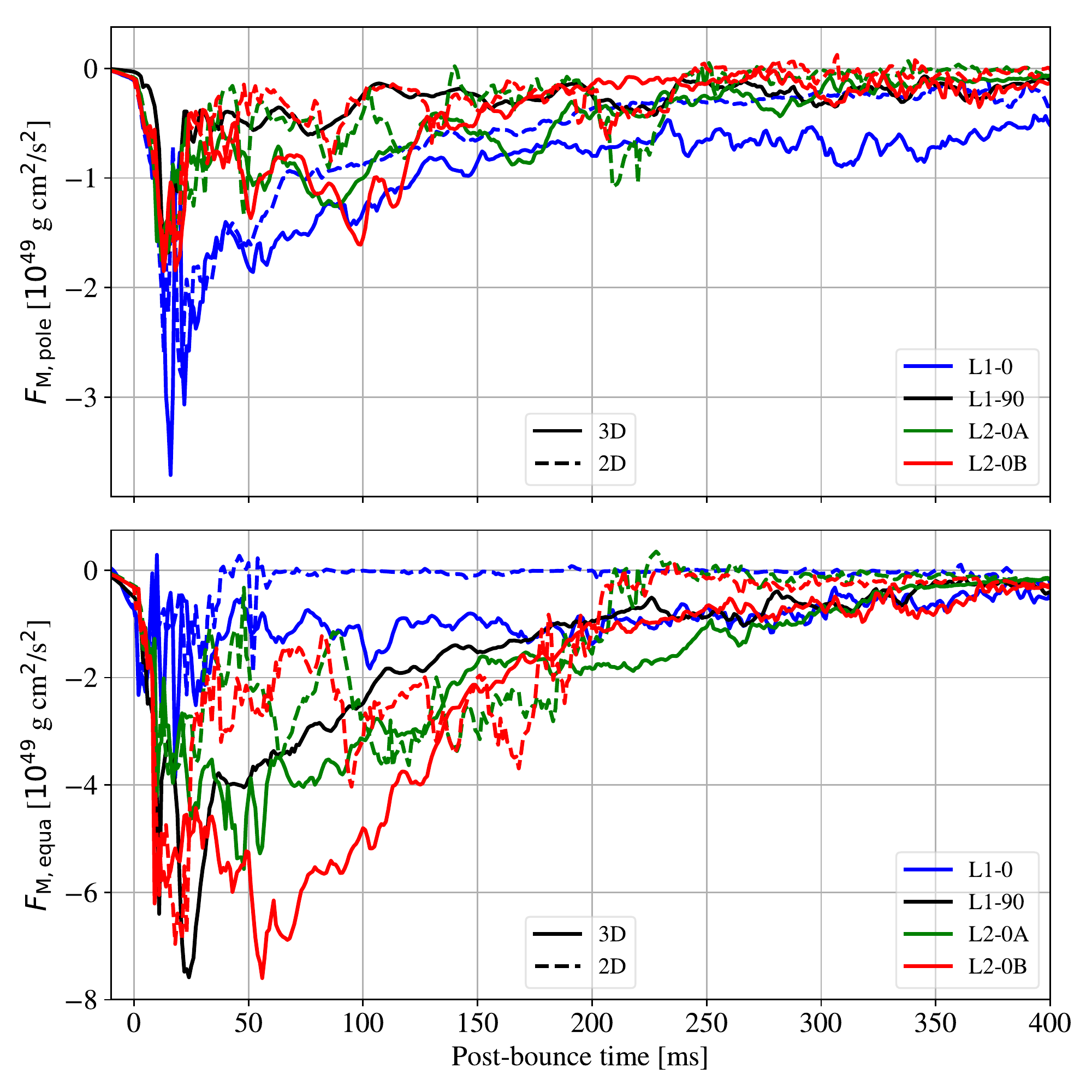}
    \caption{Time evolution of the flux of angular momentum through the PNS surface due to magnetic forces.
    The lower panel shows the flux through an equatorial region with $\pi/4<\theta<3\pi/4$, while the upper panel shows the flux through the complementary polar caps of the PNS.}
    \label{fig:angmom_flux}
\end{figure}

\subsection{Magnetic field dynamics}

Magnetic fields clearly have a deep impact on the dynamics of those models in which they are present.
An important effect is the transport of angular momentum across the PNS, extracting rotational energy from it and powering the polar outflows during the explosion.
In \refig{fig:angmom_flux} we report the time evolution of the flux of angular momentum through the surface of the PNS due to magnetic stresses, i.e.
\begin{equation}\label{eq:angmom_flux}
    F_{M,i}=\int_\mathrm{PNS,i}r\sin\theta B_\phi (\bm{B}_P\cdot \bm{n})\mathrm{~d}S.
\end{equation}
In the previous expression $\bm{B}_P$ is the poloidal component of the magnetic field and the integral on the surface of the PNS with normal vector $\bm{n}$ extends either to an equatorial region with $\pi/4<\theta<3\pi/4$ or to the complementary polar caps of the PNS.
The angular momentum flux through the equatorial region is generally larger than its polar counterpart.
Confronting the bottom panel of \refig{fig:angmom_flux} with \refig{fig:PNS} we can see a clear anti-correlation between $L_\mathrm{PNS}$ and $F_M$ in the equatorial region, showing the direct link between magnetic stresses and rotation in the PNS.
The models with a quadrupolar field and the one with a tilted dipole exhibit the strongest transport of angular momentum away from the central PNS through the regions far from the poles, confirming the qualitative impact of a non-vanishing radial magnetic field in the equatorial plane.
However, these models do not produce the most energetic explosions.
This apparent contradiction is solved once we focus once again on \refig{fig:omega_r}.
Both models \quadA{} and \quadB{} have a PNS surrounded by a fast rotating region supported by rotation, which has more angular momentum with respect to the hydrodynamic case.
This points to the fact that part of the rotational energy extracted from the PNS is not directed into the outflow, but stays instead in its proximity.
As argued in \cite{bugli2020}, the reason for this qualitative difference between dipolar and quadrupolar fields lies in the higher efficiency of the former in connecting the equatorial regions (where most of the rotational energy is stored) with the polar ones (where the jet develops due to the magnetic pressure gradients).
Such a scenario is confirmed in the top panel of \refig{fig:angmom_flux}, where model \aligneddip{} systematically shows the highest flux of angular momentum through the polar regions (solid blue line).
It is also interesting to note that 3D models have in general a more efficient transport of angular momentum than their axisymmetric counterparts, both in the polar regions (which is consistent with the more energetic ejecta associated to them) and close to the equator.
Finally, the difference in the ejecta energy between models \quadA{} and \quadB{} shown in \refig{fig:expenergy_rshock} can be better understood if we consider that the flux of angular momentum across the polar region of the PNS is higher in the former model between 150 and 300 ms p.b., which corresponds to the time interval where model \quadA{} shows an increase in the growth rate of the ejecta energy.
Such a deviation could be explained by considering the fact that the model \quadB{} produces a PNS with a shallower potential well at bounce (by roughly 30\%) and a weaker average magnetic field.

\begin{figure}
    \centering
    \includegraphics[width=0.45\textwidth]{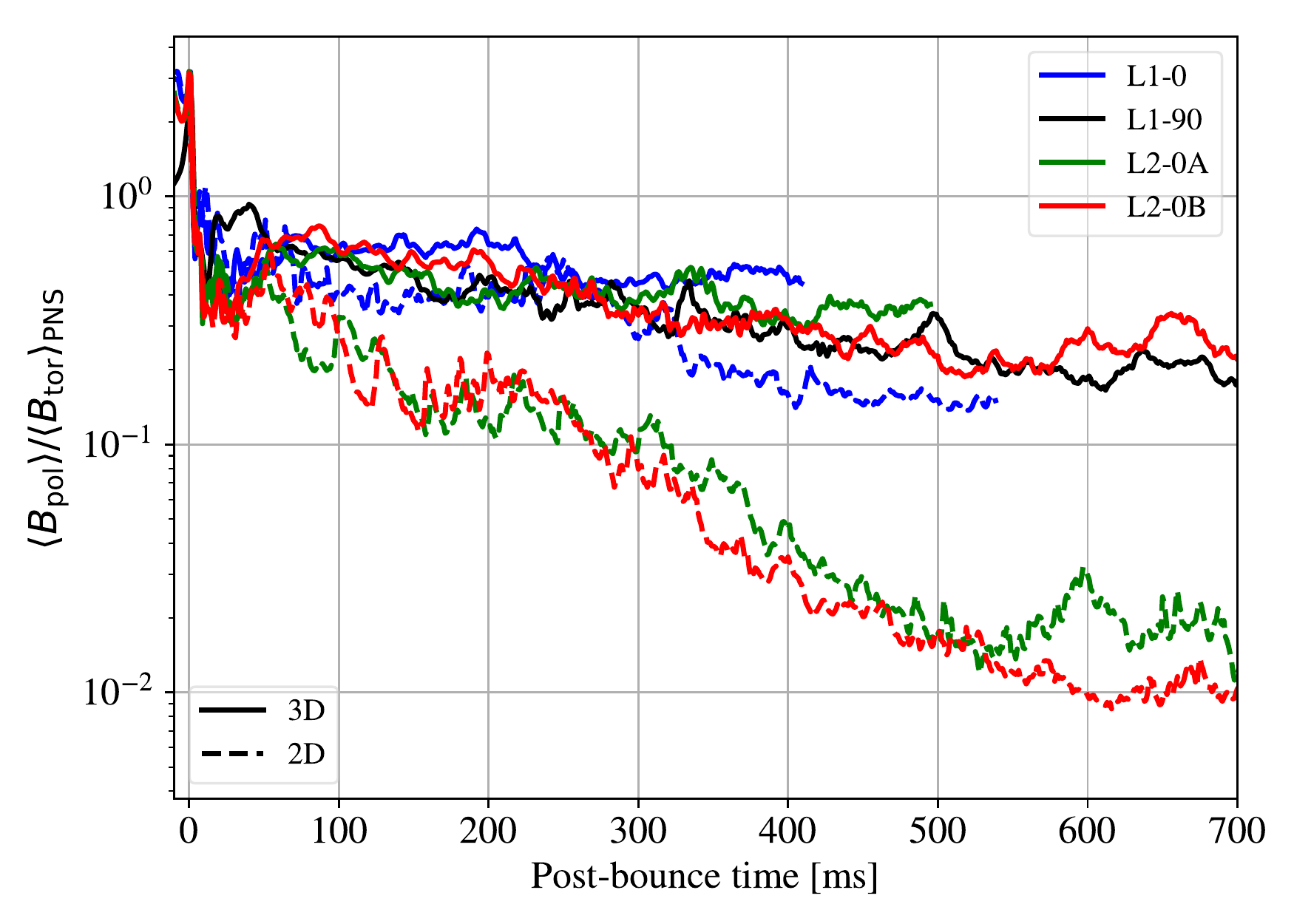}
    \caption{Time evolution of the ratio between poloidal and toroidal magnetic field averaged over the surface of the PNS.}
    \label{fig:poltor}
\end{figure}

\begin{figure}
    \centering
    \includegraphics[width=0.45\textwidth]{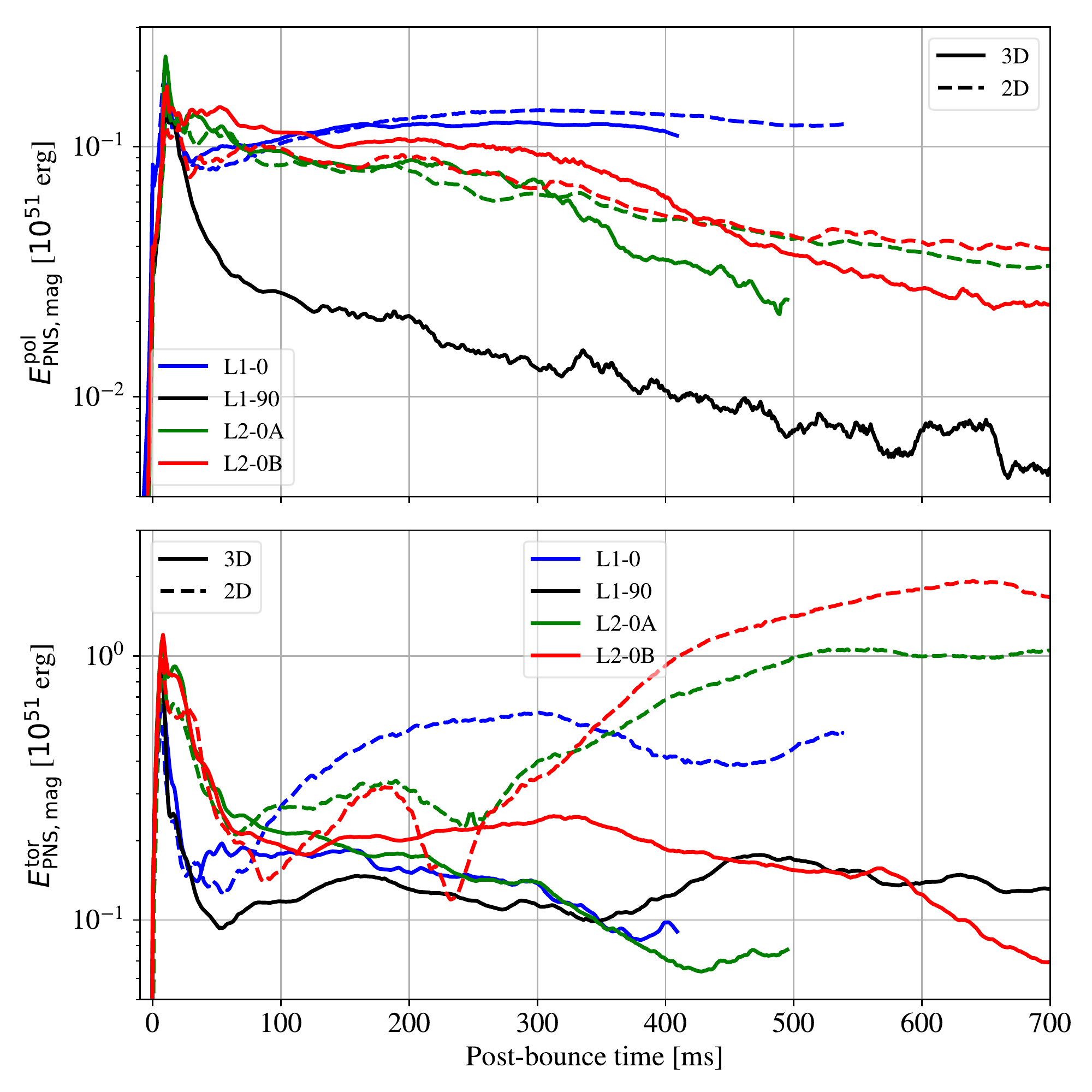}
    \caption{Time evolution of the magnetic energy within the PNS, both poloidal (top panel) and toroidal components (bottom panel).}
    \label{fig:PNS_mag_energy}
\end{figure}

Considering the ratio of poloidal to toroidal field at the PNS surface (\refig{fig:poltor}), we see a clear tendency of 3D models to produce values that are up to one order of magnitude higher than those of 2D ones (with the only exception of model \aligneddiptwod{}, for which the ratio is closer to the ones measured in 3D models).
It is tempting to interpret this feature as a consequence of the kink instability.
This instability indeed grows if the ratio of poloidal to toroidal field is low and relies on the energy stored in the toroidal magnetic field to convert it into poloidal field and kinetic energy.
The evolution of poloidal and toroidal field inside the PNS is consistent with this scenario (\refig{fig:PNS_mag_energy}).
Axisymmetric quadrupolar models experience a strong growth of the toroidal magnetic field during the whole simulation, with a clear amplification starting at $\sim$250 ms p.b.
Their 3D counterparts, instead, lack this clear growth, having a toroidal magnetic energy that remains of the order of $10^{50}$ erg.
Since the toroidal field growth is mainly due to the winding of field lines due to differential rotation (the so-called \emph{$\Omega$-effect}), the prominent growth seen in axisymmetric models is likely caused by the fact that the outward transport of angular momentum is less effective, hence the PNS retains a larger fraction of its rotational energy, leading to a more efficient amplification of the toroidal field.   
The differences in poloidal field between different models with aligned quadrupolar fields are instead more modest, with a tendency for 3D simulations to have at first a stronger poloidal component.
Moreover, at later times the poloidal field is dissipated at a faster rate than in the case of axisymmetric models, which show after $\sim400$ ms p.b. a component stronger than their 3D counterparts.
This effect could be explained by invoking the development of 3D turbulence, which determines an increase of the numerical dissipation of magnetic field at the smallest spatial scales resolved by the simulation.
% provides a mechanism to both locally amplify the magnetic field and dissipate it.
The scenarios presented by the two dipolar models are instead somewhat different.
The only difference that model \aligneddip{} shows with respect to its axisymmetric counterpart is in the toroidal field, which does not undergo the same amplification as in model \aligneddiptwod{}.
This suggests a direct link between the development of non-axisymmetric structures and the evolution of the toroidal field, and in particular with the kink instability.
Finally, model \tilteddip{} (which obviously has no 2D counterpart) shows a toroidal field consistent with the other 3D models, but also a poloidal field that within the first 50 ms p.b. decays at a much faster rate than the rest of the models.
Such a strong dissipation can be explained by considering that an equatorial dipole in a differentially rotating fluid produces a striped structure due to the winding, with the radial magnetic component inverting polarity along the radial direction.

\begin{figure}
    \centering
    \includegraphics[trim = {0 0.9cm 0 0}, clip, width=0.45\textwidth]{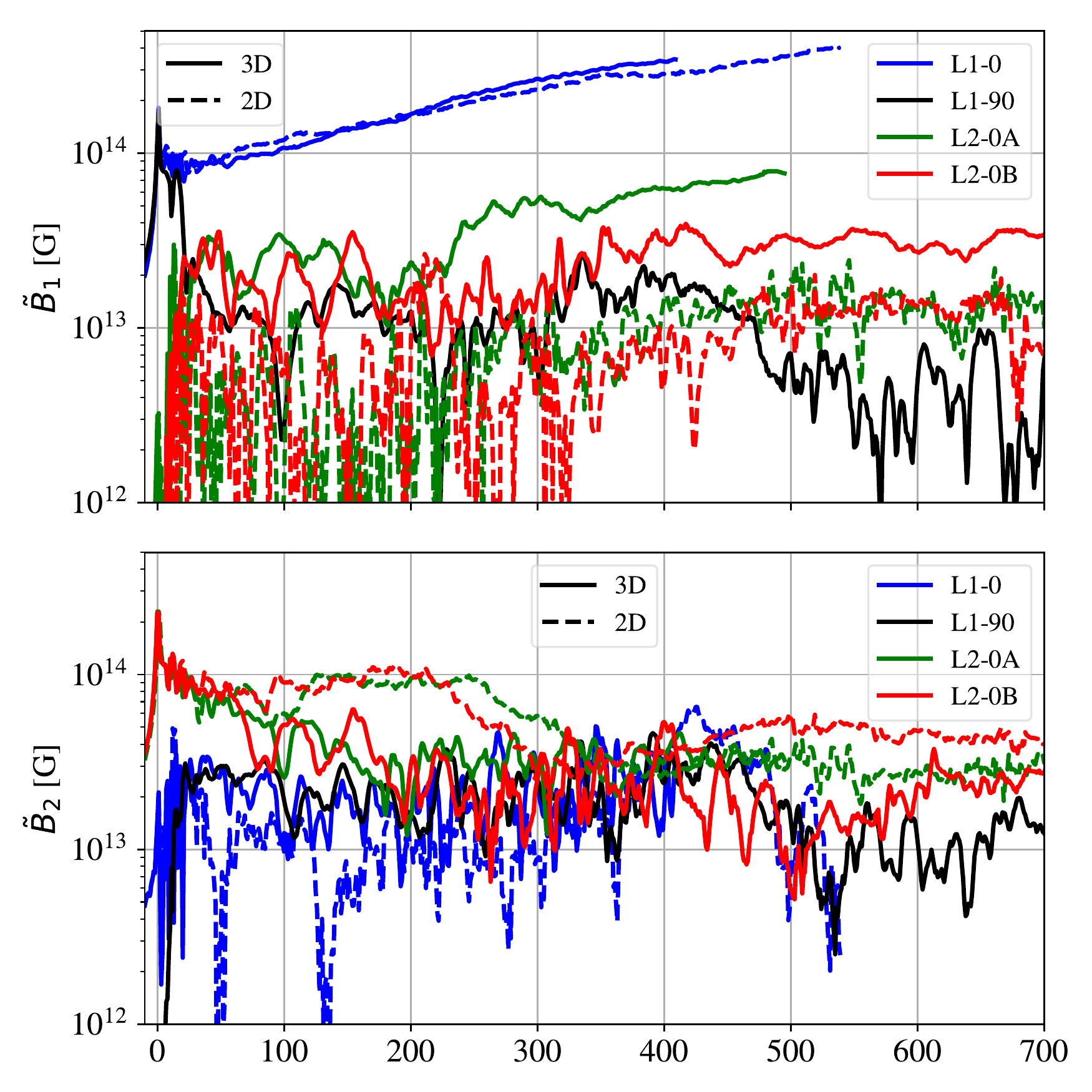}
    \includegraphics[trim = {0 0 0.25cm 0}, clip,width=0.44\textwidth]{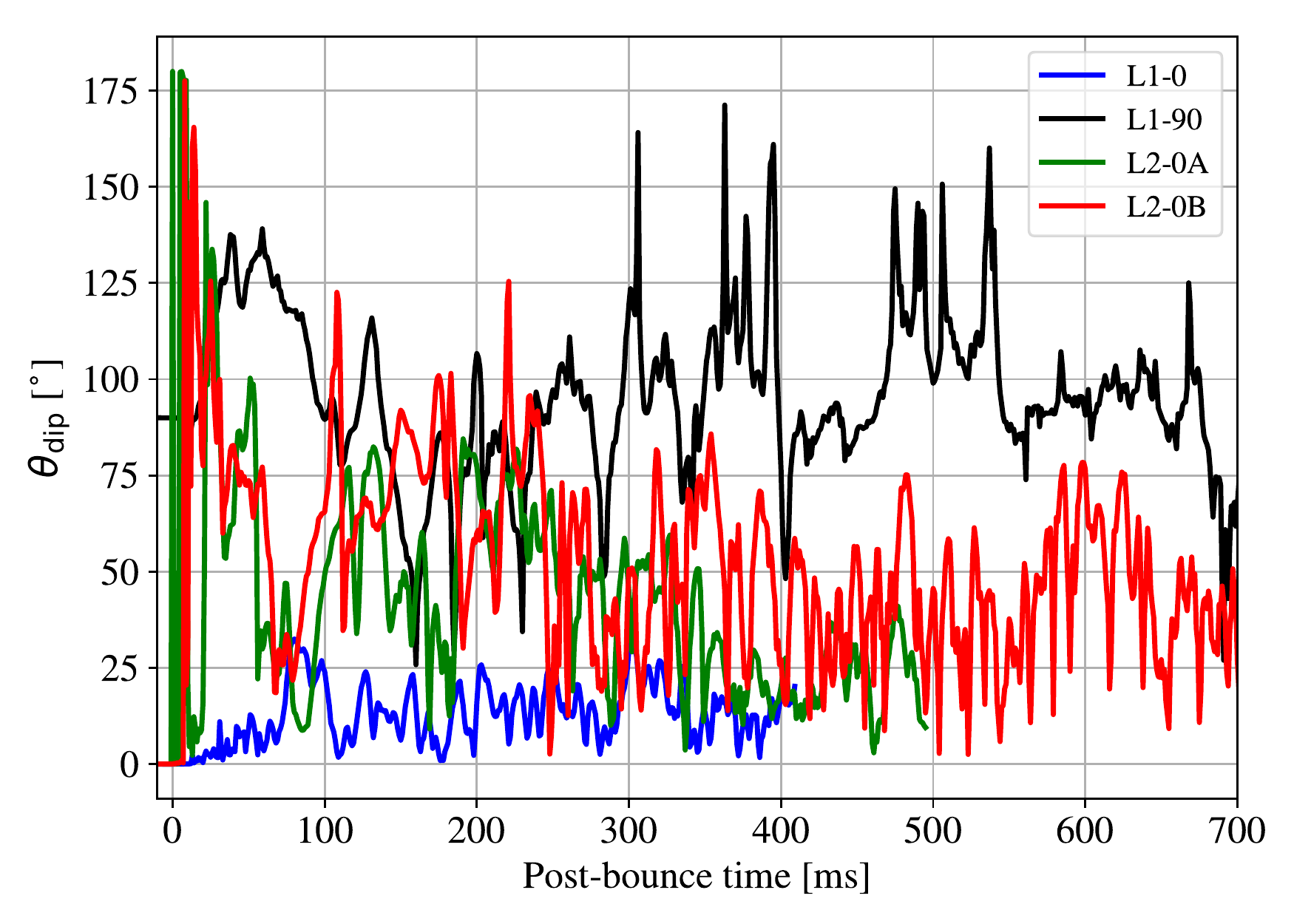}
    \caption{Time evolution of the $l=1,2$ components of the radial magnetic field (top and middle panels) as defined in \refeq{eq:Bm1} and tilt angle of the magnetic dipole at the surface of the PNS (bottom).}
    \label{fig:l1Br_tiltangle}
\end{figure}

One of the most important features of three-dimensional MHD models, as opposed to axisymmetric ones, is the ability to capture the action of dynamo mechanisms that can amplify the strength of the magnetic field beyond what can be achieved through the advection of magnetic flux.
We focus on the multipolar components of the radial magnetic field at the PNS surface, including both axial ($m=0$) and non-axisymmetric ($m=-l,\dots,l$) components, by calculating the quantity
\begin{equation}\label{eq:Bm1}
    \tilde{B}_{l}=(2l+1)\sum_{m=-l}^{l}\int \bar{B}_r(\theta,\phi) Y_{lm} \mathrm{d}\Omega.
\end{equation}
where $\bar{B}_r$ is the radial magnetic field at the PNS surface averaged over 3 cells along the radial direction.
In the top panel of \refig{fig:l1Br_tiltangle} we see a clear amplification of the dipolar field in model \aligneddip{}, which is mostly due to the advection of magnetic field onto the PNS surface and the overall contraction of the PNS.
While models with an aligned quadrupolar field start with no dipolar component, we see a clear growth of $\tilde{B}_{10}$, with model \quadB{} showing a delay of about 100 ms with respect to \quadA{}.
Such component is instead systematically weaker in our axisymmetric simulations, with large-amplitude oscillations up to 300 ms p.b. that are instead absent in the correspondent 3D models.
For model \tilteddip{} $\tilde{R}_1$ it matches the value obtained with an aligned quadrupole only up to $\sim$250 ms p.b., since it decreases significantly after that point.
Another interesting insight comes from the evolution of the quadrupolar component of the radial magnetic field (middle panel of \refig{fig:l1Br_tiltangle}), which promptly decays after bounce in the 3D models.
Such component remains instead almost constant in the first 250 ms for the axisymmetric models, but then it starts to decrease as well, reaching similar values to those produce by model \quadA{} and \quadB{}.

These results show that the topology of the magnetic field at the surface of the PNS undergoes significant changes during the gravitational collapse, with the tendency of increasing its dipolar component and reducing the quadrupolar one.
Given the relatively low resolution we adopted in this study and the associated numerical diffusion, it is difficult to clearly identify the source of the amplification of the dipolar magnetic field seen in some of our models.
Moreover, our results do not exclude the possibility that there might be some degree of stochasticity in the occurrence of the growth of the magnetic dipole, whose investigation will require the realisation of further simulations.
However, the fact that axisymmetric models produce qualitatively different evolution of the large-scale magnetic field suggests that a fully three-dimensional description of the collapse is required to capture such an effect.

We now consider the tilt-angle of the dipolar field with the rotation axis, which provides useful insights on the dynamics of the large-scale magnetic field, and we define it as 
\begin{equation}
    \theta_\mathrm{dip} = \frac{\pi}{2} - \arctan\left(\frac{\mu_z}{\sqrt{\mu_x^2+\mu_y^2}}\right),
\end{equation}
where $\mu_x$, $\mu_y$ and $\mu_z$ are the Cartesian components of the magnetic dipole moment 
\begin{equation}
    \bm{\mu}=\frac{1}{2}\int_\mathrm{PNS}\bm{r}\times\bm{J}\mathrm{d}V,
\end{equation}
where $\bm{J}$ is the electric current density and the volume integral extends to the whole PNS.
In the bottom panel of \refig{fig:l1Br_tiltangle} we can see that the initially aligned and perpendicular dipoles approximately keep their initial orientation on average, while models \quadA{} and \quadB{} assume intermediate tilt angles.
Note, however, that the dipolar field in model \aligneddip{} does not remain strictly aligned to the rotation axis, but develops instead a small tilt angle of $\sim10^{\circ}$ which appears to slowly grow.
This effect explains why, despite model \aligneddiptwod{} presenting a negligible transport of angular momentum in the equatorial region (\refig{fig:angmom_flux}), for its 3D counterpart $F_M$ reaches significant values that become comparable with the ones produced by the quadrupolar models after 200 ms p.b.: a small tilt in the magnetic dipole means that a non-vanishing radial field is present in the equatorial region, hence the magnetic transport of angular momentum becomes more efficient in that direction.
The tilt angle of the equatorial dipole of model \tilteddip{} alternates instead periods of temporary growth or decrease with quick oscillations around $90^{\circ}$, with amplitudes up to $\sim60^{\circ}$.
This is consistent with the lack of a significant axial dipole seen in the top panel of \refig{fig:l1Br_tiltangle} for the model with tilted dipolar field.
The dipolar component developed by models \quadA{} and \quadB{} starts instead with very fast and large amplitude oscillations across the equator in the first few tens of ms, then the tilt angle decreases in mean value and approaches values in between $30^\circ-40^\circ$.

These results seem to point to the fact that an initial magnetic field that differs from an aligned dipole (either by its inclination or topology) tend to produce on the surface of the PNS dipolar fields with are not aligned to the rotational axis.
Such a scenario has been obtained by recent numerical models of local dynamos within the PNS \citep{raynaud2020,reboul-salze2021}, where the action of convection and the MRI lead to the formation of large-scale fields whose dipolar component can exhibit a very large inclination.

%%%%%%%%%%%%%%%%%%%%%%%%%%%
%%%%%%%%%%%%%%%%%%%%%%%%%%%
%%%%%%%%%%%%%%%%%%%%%%%%%%%
%%%%%%%%%%%%%%%%%%%%%%%%%%%
%%%%%%%%%%%%%%%%%%%%%%%%%%%
%%%%%%%%%%%%%%%%%%%%%%%%%%%
%%%%%%%%%%%%%%%%%%%%%%%%%%%
%%%%%%%%%%%%%%%%%%%%%%%%%%%
%%%%%%%%%%%%%%%%%%%%%%%%%%%
%%%%%%%%%%%%%%%%%%%%%%%%%%%
%%%%%%%%%%%%%%%%%%%%%%%%%%%
%%%%%%%%%%%%%%%%%%%%%%%%%%%
%%%%%%%%%%%%%%%%%%%%%%%%%%%
%%%%%%%%%%%%%%%%%%%%%%%%%%%
%%%%%%%%%%%%%%%%%%%%%%%%%%%
%%%%%%%%%%%%%%%%%%%%%%%%%%%
%%%%%%%%%%%%%%%%%%%%%%%%%%%
%%%%%%%%%%%%%%%%%%%%%%%%%%%

\section{Conclusions}\label{sec:conclusions}
We presented a series of 3D relativistic MHD simulations of CCSN of the same fast rotating stellar progenitor with different initial magnetic field configurations departing from the simple case of an aligned magnetic dipole, as such assumption is unlikely to reflect the real complexity that the field topology can assume.
Our models included an unmagnetised case and one with an aligned dipole as benchmarks to assess the effects of the more complex magnetic field topologies we explored, i.e. aligned quadrupolar fields and equatorial dipoles.

Although all of our models produce successful explosions, the aligned dipole case produces the most energetic ejecta and the fastest shock expansions, while the quadrupolar fields and equatorial dipole lead to progressively weaker explosions (but still more energetic than the hydrodynamic model).
The morphology of the ejecta differs significantly among different models, with \aligneddip{} producing a well collimated and almost axisymmetric bipolar outflow.
For the other magnetised models the ejecta still expand preferentially along the rotation axis, although in case of the equatorial dipole the shock is rather spherical.

The PNSs produced by the magnetic models end up having similar masses, which are smaller than both the hydrodynamic case and the corresponding axisymmetric models.
The rotation profile, on the other hand, evolves in very different ways depending on the magnetic field configuration.
Models with quadrupolar fields or an equatorial dipole exhibit a more efficient spin down of the PNS, due to a more efficient transport of angular momentum in the equatorial region caused by a non-vanishing radial field.
However, this effect does not result in an enhancement of the magnetorotational explosion mechanism, as the extracted rotational energy is then deposited in the surroundings of the PNS equatorial plane, rather than close to the polar caps where the ejection is launched.
On the other hand, with an aligned dipolar field the transport of angular momentum from the polar regions of the PNS is much more efficient, leading to an overall more energetic explosion despite leaving a faster rotating PNS.

All our magnetised 3D models showed the development of the kink instability around the time of core bounce, with a clear exponential growth during a short linear phase and a quick saturation that lasts for the rest of the simulation,
corroborating the results of \cite{obergaulinger2021a} and extending them to the case of different magnetic topologies.
The linear growth rate of the instability we measured is of the same order of magnitude as the one found by \cite{mosta2014b}, showing a certain degree of consistency between this study and ours.
However, the model they considered did not produce a powerful prompt explosion with a well collimated jet, despite employing an aligned dipolar field (but considering a different progenitor and distribution of angular momentum with respect to the present work).
Thus, it is still not clear to what extent the kink instability might still affect the launch of magnetorotational explosions for a generic choice of initial conditions (such as rotation profile, magnetic field strength and distribution between poloidal and toroidal components).
Moreover, it is also not understood what might be the impact on the development of the kink instability of the specific numerical methods employed in numerical codes, such as, for example, the topology of the numerical grid (Cartesian or spherical coordinates).
More studies will be required to quantitatively address this problem, possibly comparing the results obtained by different codes using the same set of initial conditions.

Although our results show a qualitative agreement in the explosion dynamics between axisymmetric and three-dimensional models, there are some systematic differences that lead to significant deviations.
The energy of the unbound ejecta and the velocity of the outgoing shock in the 3D models are higher than those produced in their axisymmetric counterparts, just as the transport of angular momentum from the PNS.
This is related to the different evolution of the axial dipolar component of the magnetic field, which increases more in the 3D case and thus allows a more efficient extraction of the rotational energy from the PNS.
Such a growth of the large-scale dipolar component is instead quenched in 2D models, where no sustained dynamo mechanism can occur.
The different amplification of the magnetic field has a direct impact on the magnetic transport of angular momentum from the PNS, which is systematically more effective in 3D models than in their 2D counterparts.
Overall these results are consistent with the findings of \cite{obergaulinger2021a}, which showed that relaxing the assumption of axisymmetry tend to produce more energetic magnetorotational explosions and faster expanding shocks.
Moreover, we showed that taking into account the full geometry of the problem decreases the growth of the toroidal field in the PNS due to differential rotation, which is likely connected to the development of the kink instability and a more efficient extraction of angular momentum from the PNS.

The study of the dynamical impact of magnetic configurations such as quadrupoles and equatorial dipoles is justified by the recent findings that dynamo mechanisms occurring within the PNS and driven by either convection \citep{raynaud2020,raynaud2021} or the magnetorotational instability \citep{reboul-salze2021} tend to produce large-scale magnetic fields that are rarely aligned dipoles, but shows instead significant inclinations with respect to the rotation axis and a distribution on higher multipoles.
Although our models could qualitatively capture the effects of more complex magnetic structures, they clearly lacked the resolution required to self-consistently reproduce the amplification of the magnetic field in the PNS and trace a more robust connection between PNS-driven dynamos and magnetorotational explosions.
A possible solution to this problem could be the use of subgrid models to include a mean-field dynamo mechanism in core-collapse simulations, which could be calibrated by local high-resolution models of the PNS internal regions.

%%%%%%%%%%%%%%%%%%%%%%%%%%%
%%%%%%%%%%%%%%%%%%%%%%%%%%%
%%%%%%%%%%%%%%%%%%%%%%%%%%%
%%%%%%%%%%%%%%%%%%%%%%%%%%%
%%%%%%%%%%%%%%%%%%%%%%%%%%%
%%%%%%%%%%%%%%%%%%%%%%%%%%%
%%%%%%%%%%%%%%%%%%%%%%%%%%%
%%%%%%%%%%%%%%%%%%%%%%%%%%%
%%%%%%%%%%%%%%%%%%%%%%%%%%%
%%%%%%%%%%%%%%%%%%%%%%%%%%%
%%%%%%%%%%%%%%%%%%%%%%%%%%%
%%%%%%%%%%%%%%%%%%%%%%%%%%%
%%%%%%%%%%%%%%%%%%%%%%%%%%%
%%%%%%%%%%%%%%%%%%%%%%%%%%%
%%%%%%%%%%%%%%%%%%%%%%%%%%%
%%%%%%%%%%%%%%%%%%%%%%%%%%%
%%%%%%%%%%%%%%%%%%%%%%%%%%%
%%%%%%%%%%%%%%%%%%%%%%%%%%%

\section*{Acknowledgements}

MB and JG acknowledge support from the European Research Council (ERC starting grant no. 715368 -- MagBURST) and from the \emph{Tr\`es Grand Centre de calcul du CEA} (TGCC) and GENCI for providing computational time on the machines IRENE and OCCIGEN (allocation A0050410317).
MO acknowledges support from the Spanish Ministry
of Science, Education and Universities (PGC2018-095984-B-I00)
and the Valencian Community (PROMETEU/2019/071), the European Research Council under
grant EUROPIUM-667912, and from the the Deutsche Forschungsgemeinschaft
(DFG, German Research Foundation) – Projektnummer
279384907 – SFB 1245 as well as from the Spanish Ministry
of Science via the Ramón y Cajal programme (RYC2018-024938-I).
MO acknowledges as well the support through the grants AYA2015-66899-C1-1-P and PROMETEOII-2014-069 of the Spanish Ministry of Economy and Competitiveness (MINECO) and of the Generalitat Valenciana, respectively.
JG acknowledges the support from the PHAROS COST Action CA16214.

\section*{Data availability}
The data underlying this article will be shared on reasonable request to the corresponding author.
%%%%%%%%%%%%%%%%%%%%%%%%%%%%%%%%%%%%%%%%%%%%%%%%%%



\bibliographystyle{mnras}
\bibliography{references} 

%%%%%%%%%%%%%%%%%%%%%%%%%%%%%%%%%%%%%%%%%%%%%%%%%%

%%%%%%%%%%%%%%%%% APPENDICES %%%%%%%%%%%%%%%%%%%%%

% \appendix

%%%%%%%%%%%%%%%%%%%%%%%%%%%%%%%%%%%%%%%%%%%%%%%%%%

% Don't change these lines
\bsp	% typesetting comment
\label{lastpage}
\end{document}